\documentclass[aps, 10pt, amsmath, twocolumn, superscriptaddress]{revtex4-1}
\usepackage[plainpages=false,colorlinks=true,linkcolor=blue, citecolor=blue, urlcolor=blue]{hyperref}
\usepackage[english]{babel}
\usepackage{graphicx}
\usepackage[charter]{mathdesign}
\usepackage{bm}
\usepackage[svgnames]{xcolor}
\usepackage{subfigure}
\usepackage{booktabs}

\usepackage{array}

\newcolumntype{C}{>{$\displaystyle}c<{$}}
\newcolumntype{L}{>{$\displaystyle}l<{$}}
\newcolumntype{R}{>{$\displaystyle}r<{$}}
\newcolumntype{A}{ >{$} r <{$} @{\extracolsep{0pt}} >{${}} l <{$} }
\usepackage{array}

\newcommand\rv{\mathbf{r}}
\newcommand\xv{\mathbf{x}}
\newcommand\yv{\mathbf{y}}

\begin{document}

\title{Doping dependence of chiral superconductivity in near $45^\circ$ twisted bilayer cuprates}

\author{Mathieu B\'elanger}
\affiliation{D\'epartement de physique and Institut quantique, Universit\'e de Sherbrooke, Sherbrooke, Qu\'ebec, Canada J1K 2R1}
\author{David S\'en\'echal}
\affiliation{D\'epartement de physique and Institut quantique, Universit\'e de Sherbrooke, Sherbrooke, Qu\'ebec, Canada J1K 2R1} 
\date{\today}

\begin{abstract}
We study a one-band Hubbard model for a twisted cuprate bilayer with a twist angle of 43.6$^\circ$ and a moire cell containing 58 sites. We use the variational cluster approximation (VCA), which treats short-range correlations exactly and leads, in single layers, to a dome of $d$-wave superconductivity away from half-filling from strong on-site repulsion alone. We find a time-reversal-symmetry (TRS) breaking phase in a small doping interval in the overdoped region when inter-layer tunneling is strong enough. Contrary to expectations, being closer to the $45^{\circ}$ twist angle does not expand this TRS region compared to a previous study~[Lu \textit{et al}, Phys. Rev. B \textbf{105}, 245127 (2022)] on a $53^{\circ}$ twist angle. This is attributed to the fact that the two superconducting states in competition have almost identical nodal structures.
\end{abstract}
%\pacs{}
\maketitle

%===============================================================================
\section{Introduction}

The observation of unconventional superconductivity in twisted bilayer graphene~\cite{cao_correlated_2018,cao_unconventional_2018} has motivated similar studies on various van der Waals heterostructures~\cite{balents_superconductivity_2020}, like twisted boron nitride~\cite{xian_multiflat_2019} and transition metal dichalcogenides~\cite{ruiz-tijerina_interlayer_2019,wang_correlated_2020,an_interaction_2020,zhang_flat_2020,venkateswarlu_electronic_2020,regan_mott_2020,tang_simulation_2020,belanger_superconductivity_2022}. Those systems offer more degrees of freedom than their monolayer counterparts due to the twist angle that can be changed.

The realization of a two-dimensional monolayer of Bi${}_2$Sr${}_2$CaCu${}_2$O${}_{8+\delta}$ (Bi2212)~\cite{yu_high-temperature_2019,zhao_sign-reversing_2019} has motivated research on twistronics in cuprates~\cite{can_high-temperature_2021,volkov_magic_2023,lu_doping_2022,margalit_chiral_2022,mercado_high-temperature_2022}. It has been predicted that a fully gapped, time-reversal symmetry (TRS) breaking $d_{x^2-y^2}+id_{xy}$ superconducting phase emerges in systems with twist angles close to $45^\circ$~\cite{can_high-temperature_2021}. This superconducting phase is predicted to be topologically nontrivial. Chiral topological superconductivity is also predicted in twisted multilayer nodal superconductors~\cite{tummuru_twisted_2022}.

The bilayer cuprate systems are natural Josephson junctions. Realizing those junctions in the laboratory is challenging due to disorder that may be introduced while preparing the sample~\cite{martini_twisted_2023,lee_encapsulating_2023}. An early study of a c-axis twisted cuprate Josephson junction did not show the expected $d$-wave behavior~\cite{zhu_presence_2021}. New processes have been proposed to create those junctions and offer evidence that the Josephson current is reduced close to $45^\circ$ due to a mismatch between the $d$-wave states of the two layers~\cite{lee_twisted_2021,martini_twisted_2023,lee_encapsulating_2023}. The critical current might remain finite at $45^\circ$, pointing to a state with TRS breaking~\cite{zhao_time-reversal_2023}. Such behavior has been predicted using simple models describing twisted bilayer cuprates~\cite{volkov_josephson_2021,tummuru_josephson_2022}.

Previous theoretical work based on Bogoliubov-de-Gennes mean-field theory lacks the effect of strong correlation, which are important in cuprates. Moreover, the physics of cuprates is strongly doping dependent, being affected by the pseudogap phenomenon below optimal doping~\cite{dash_pseudogap_2019}. A twisted $t-J$ model of cuprates within slave-boson mean field theory predicts that the range of twist angle around $45^\circ$ which allows TRS breaking is narrow~\cite{song_doping_2022}. To address the effect of the strong correlation and the effect of doping, a Hubbard model for a system with a twist angle of $\theta=53.13^\circ$ has been studied with the variational cluster approximation (VCA) and cluster dynamical mean field theory (CDMFT)~\cite{lu_doping_2022}. The TRS breaking was observed near optimal doping when strong inter-layer tunneling was considered. It was conjectured that at a twist angle closer to $45^\circ$, the mixed state could be more stable as a function of doping. The number of orbitals per unit cell in systems close to $45^\circ$ makes them challenging to study using quantum cluster methods.

In this paper we take a step further into studying systems close to $45^\circ$, by considering a Hubbard model for bilayer cuprates at a twist angle $\theta=43.60^\circ$, corresponding to a unit cell of 58 copper sites. 
We use the VCA to probe the stability of the TRS-breaking superconducting phase against doping. 
We follow the methodology proposed in Ref.~\cite{lu_doping_2022} and show that a TRS-breaking phase occurs again in a narrow doping range, which can be explained by the band structure limiting the energy gained from combining the two superconducting states in competition. 
Thus, moving closer to a $45^\circ$ twist angle does not necessarily stabilizes the TRS-breaking phase, which is counter-intuitive.

%===============================================================================
\section{Model and Method}
\label{sec:model}

\subsection{Model}

We use the tight-binding Hubbard model proposed in Ref.~\cite{lu_doping_2022}, where each layer is described by a one-band Hubbard model, each site corresponding to a copper atom. The Hamiltonian can be separated as 
\begin{equation}
    H=H^{(1)}+H^{(2)}+H_\perp,
    \label{eq:H}
\end{equation}
where the intralayer Hamiltonian $H^{(l)}$ is
\begin{equation}
    H^{(l)}=\sum_{\rv,\rv'\in l,\sigma}t_{\rv\rv'}c^\dagger_{\rv,l,\sigma}c_{\rv',l,\sigma}+U\sum_{\rv}n_{\rv,l,\uparrow}n_{\rv,l,\downarrow}-\mu\sum_{\rv,\sigma}n_{\rv,l,\sigma}
    \label{eq:H-intra}
\end{equation}
where $c_{\rv,l,\sigma}$ $(c^\dagger_{\rv,l,\sigma})$ is the annihilation (creation) operator of an electron at site $\rv$ on layer $l$ ($l=1,2$) with spin $\sigma=\uparrow,\downarrow$, and $n_{\rv,l,\sigma}$ is the number operator. $\rv,\rv'$ are the site indices of a square lattice for each layer. $U$ is the on-site repulsion between electrons and $t_{\rv\rv'}$ is the hopping matrix that includes nearest-neighbor hopping ($t$) and next-nearest-neighbor hopping ($t'$).
No extended interaction terms were considered since strongly-correlated superconductivity can be driven by local repulsion alone and has been shown to be resilient to a extended Coulomb repulsion at intermediate to strong coupling~\cite{senechal_resilience_2013}.
In Bi2212, the nearest-neighbor hopping is $t=126$ meV~\cite{markiewicz_one-band_2005}, but here we are using $t$ as the energy unit. To describe Bi2212 we use the set of parameters $t=1$, $t'=-0.3$ and $U=8$~\cite{markiewicz_one-band_2005,lu_doping_2022}. We also tested other parameter sets (Sect.~\ref{sec:other_params}) to assess the robustness of our results.

The effect of the twist angle comes from the inter-layer tunneling given by
\begin{equation}
    H_\perp=\sum_{n=1}^{7}V_n\sum_{\langle\rv,\rv'\rangle_{\perp,n,\sigma}}\left[c^\dagger_{\rv,1,\sigma}c_{\rv',2,\sigma}+\mathrm{H.c.}\right],
    \label{eq:H-inter}
\end{equation}
with $\langle\rv,\rv'\rangle_{\perp,n,\sigma}$ representing the set of sites $\rv$ on layer 1 and $\rv'$ on layer 2 such that their projection on the plane are $n^{\rm th}$ neighbors.
We will explain below why we consider inter-layer hopping up to 7$^{\rm th}$ inter-layer neighbors.
The strength of the tunneling is given by 
\begin{equation}\label{eq:expV}
    V_n=Ve^{-\lambda\left(|\mathbf{d_n}|-d_z\right)/a}, 
\end{equation} 
where $|\mathbf{d_n}|=|\rv-\rv'|$ is the three-dimensional distance between the two sites corresponding to the $n$th neighbors on different layers, $d_z$ is the distance between the two layers and $a$ is the lattice constant of the square lattice. $V$ is the inter-layer tunneling of sites that are on top of each other. $\lambda$ is a damping parameter. To have a set of parameters similar to the one used in Ref.~\cite{lu_doping_2022}, we chose $d_z=a$ and $\lambda=11.13$ with two different values of $V$: $0.2$ and $0.4$. 

This model is a simplification of the real situation in bilayer cuprates. The effect of the rare-earth layer in between the CuO${}_2$ plane is doubtless more complex. The model also neglects the fact that each monolayer of Bi2212 contains two CuO${}_2$ planes. The value of the inter-layer tunneling parameters chosen here are very strong and should be lower in order to better describe real twisted bilayers. 
Indeed, LDA calculations~\cite{markiewicz_one-band_2005} for bulk Bi2212 point towards an interlayer hopping $t_z/t\approx 0.1$, smaller than our $V=0.2$ or $0.4$. 
These strong values of $V$ are needed to observe TRS breaking in our system~\cite{lu_doping_2022}.
For inter-layer hopping between sites that are not directly superimposed, adopting the exponential model~\eqref{eq:expV} is both practical and reasonable since it allows comparisons between different twist angles while keeping the hopping parameters fixed.
Overall, the effective model used here allows us to roughly delimit the range of parameters needed for chiral superconductivity to exist and to conclude whether that phase is within or beyond the reach of experiments on Bi2212 bilayers.

In this work we consider a system with a twist angle $\theta = 43.60^\circ$, in order to probe TRS breaking close to the optimal twist angle. 
Only a few twist angles correspond to commensurate systems with a reasonably small unit cell allowing for a microscopic description based on the Hubbard model~\cite{can_high-temperature_2021}.
This drastically limits the twist angles that can be studied.
In Ref.~\cite{lu_doping_2022}, a ten-site unit cell with a twist angle of $\theta=53.13^\circ$ was used. 
The next simplest case is the one studied here, a 58-site unit cell with a twist angle of $\theta = 43.60^\circ$.
The next simpler case would correspond to a twist angle of $\theta=45.24^\circ$ with 338 sites in the unit cell, quite beyond our computing capacity.
Our goal is not to investigate all possible twist angles, but to see whether inching closer to a $45^\circ$ twist changes the doping range over which
a TRS-breaking superconducting state can be found.

%-------------------------------------------------------------------------------
\subsection{Method}

To probe the superconducting state in this model, we use the VCA~\cite{potthoff_variational_2003,dahnken_variational_2004} with an exact diagonalization solver at zero temperature. 
This is a variational approach on the electron self-energy which allows broken symmetries while treating short-range dynamical correlations exactly. 
It has been applied to magnetic phases~\cite{dahnken_variational_2004, sahebsara_hubbard_2008,laubach2014} and superconductivity~\cite{senechal2005,faye_interplay_2017,belanger_superconductivity_2022} in various systems. 
For a detailed review of the method, see Refs~\cite{potthoff_variational_2012,potthoff_dmft_2014}.
The VCA is based on a cluster decomposition of the system: the infinite lattice is tiled with identical clusters small enough for the
model to be solved exactly on each of them.
Each cluster's Hamiltonian is basically a restriction of the lattice Hamiltonian to the cluster, plus one or more \textit{Weiss fields},
whose role is to approximately represent the effect of the cluster's environment (the rest of the lattice) on the cluster's self-energy.
The VCA proceeds by optimizing the Potthoff self-energy functional as a function of these Weiss fields.
The optimal self-energy is subsequently applied to the whole lattice, and this provides an expression for the Green function of the lattice problem, which can be used to compute the average of any one-body or anomalous operator.
We use a restricted Nambu formalism to describe superconductivity, in which a particle-hole transformation was applied to the spin-down operators $c^\dagger_{\rv,l,\downarrow}$, allowing hopping terms and pairing terms to be treated on the same footing.

\begin{figure}[h]
	\begin{center}
	\includegraphics[width=0.9\hsize]{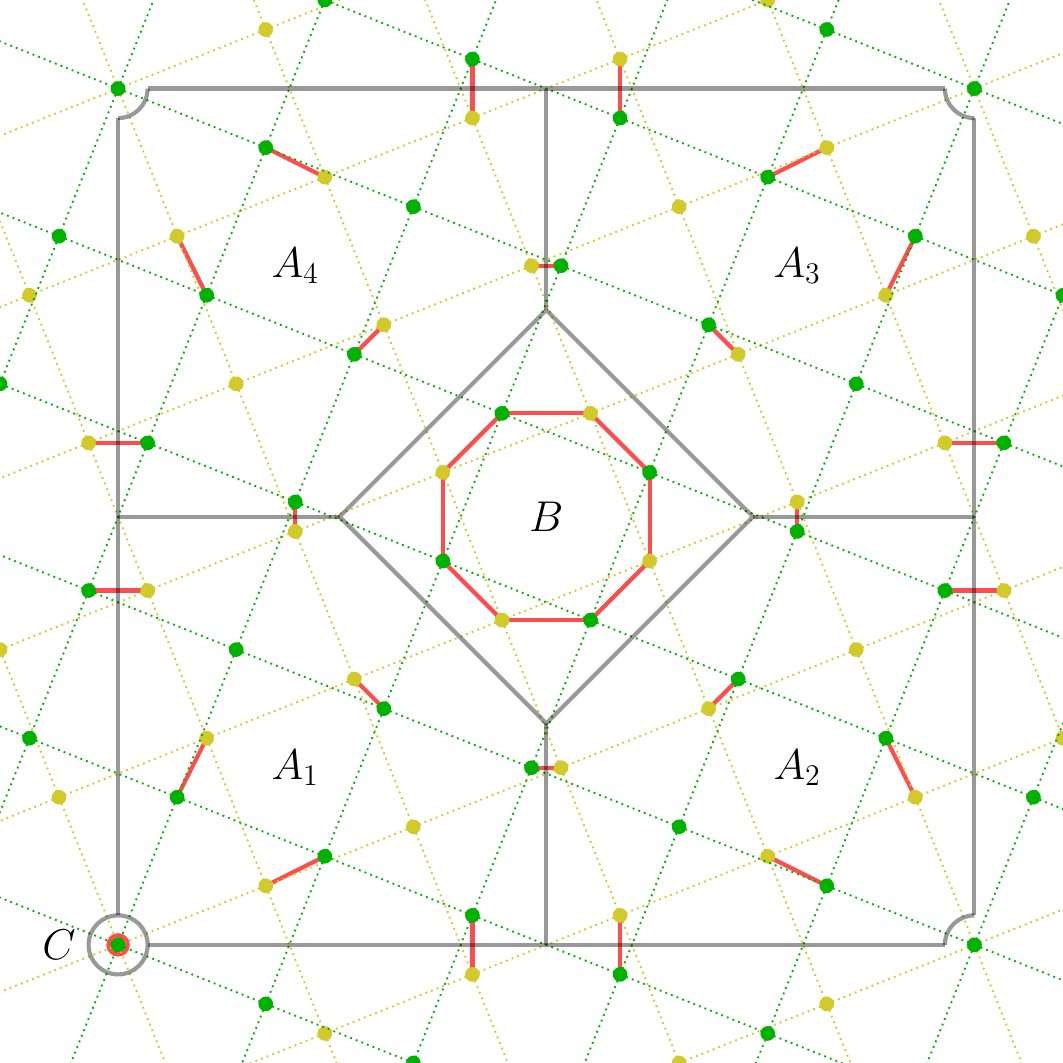}
	\caption{Unit cell of the twisted cuprate system at $\theta=43.60^\circ$. It contains 58 sites between the two layers (green and yellow). The four $A_i$ clusters contain 12 sites (6 from each layer), the $B$ cluster contains 8 sites and the $C$ cluster contains only 2 sites that are on top of each other.
	The interlayer hopping amplitudes kept in our analysis are shown in red. The largest one ($V$) occurs between superimposed sites (cluster $C$). The interlayer amplitudes in cluster $B$ are worth $0.24V$ (shorter bond) and $0.2V$ (longer bond) respectively, with our choice of $\lambda$ (see Eq.~\eqref{eq:expV})).
	}
	\label{fig:schema_58sites}
\end{center}
\end{figure}

The use of exact diagonalization limits the total number of orbitals in a cluster to about 12 because of the computational resources needed to compute the Green function. 
Considering this, we need to separate the 58-site unit cell in clusters of smaller sizes. 
We partitioned the unit cell into six clusters, as shown in Fig.~\ref{fig:schema_58sites}: one cluster of two sites (labeled $C$), one cluster of eight sites (labeled $B$) and four clusters of twelve sites each (labeled $A$).
The basic approximation of cluster methods is to neglect the components $\Sigma_{ij}(\omega)$ of the self-energy matrix for which sites $i$ and $j$ belong to different clusters.
The self-energy matrix $\mathbf{\Sigma}(\omega)$ is thus block-diagonal, each bock being associated with a cluster.
In the partition chosen here, blocks associated with clusters $A$, $B$ and $C$ would be of size 24, 16 and 4, respectively (there is a factor of 2 compared to cluster size because of the spin/Nambu degree of freedom).

The specific partition illustrated in Fig.~\ref{fig:schema_58sites} is not the only one possible, but it is one where every cluster but $C$ contains full $2\times2$ plaquettes in each layer. 
A successful description of strongly-correlated superconductivity in the single-layer Hubbard model with cluster methods requires minimally a $2\times2$ plaquette~\cite{maier_systematic_2005, senechal2005}.
Other partitions without this property were tried but were problematic within the VCA, for instance leading to discontinuities in the Potthoff functional, etc.

We keep inter-layer hopping up to 7$^{\rm th}$ inter-layer neighbors in order to ensure that every cluster contains inter-layer hopping. 
This also allows comparison with Ref.~\cite{lu_doping_2022}. 
Note that the corresponding lateral physical distances are still smaller than the nearest-neighbor distance within each layer.
For sake of illustration, the exponential hopping model \eqref{eq:expV} leads to an interlayer hopping amplitude between adjacent sites of cluster B in the range $0.2V$ to $0.24V$.

%...............................................................................
%...............................................................................

%-------------------------------------------------------------------------------
\subsection{Superconducting order parameter}

\begin{figure}[h]
	\begin{center}
	\includegraphics[width=\hsize]{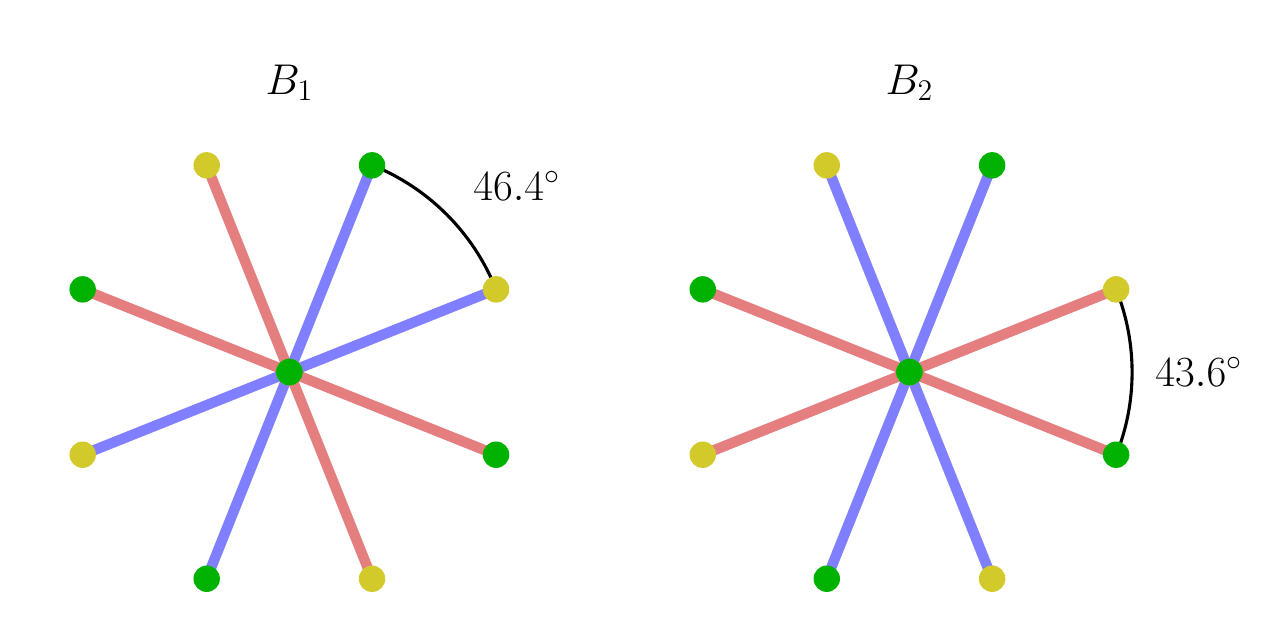}
	\caption{Schematics of the pairing operators associated with the $B_1$ and $B_2$ representations around the pivotal sites ($C$ cluster from Fig.~\ref{fig:schema_58sites}). The green (yellow) sites correspond to the top (bottom) layer. Positive (negative) pairing is represented by blue (pink) segments. In the $B_1$ representation, the links with the same sign are separated by the larger angle ($46.4^\circ$). In the $B_2$ representation, they are separated by the smaller angle ($43.6^\circ$).
	}
	\label{fig:B1B2}
\end{center}
\end{figure}

The superconducting gap in single layer or bulk cuprates has $d_{x^2-y^2}$ symmetry.
This is well understood in theoretical studies of the doped, square-lattice Hubbard model,
using many theoretical methods, including VCA~\cite{senechal2005}.
A minimal, nearest-neighbor $d$-wave pairing operator on layer $l$ is defined as
\begin{equation}
\begin{aligned}
    \hat\Delta^{(l)}=&\sum_{\rv\in l}c_{\rv,l,\uparrow}c_{\rv+\xv^{(l)},l,\downarrow}-c_{\rv,l,\downarrow}c_{\rv+\xv^{(l)},l,\uparrow}\\
    &\quad- c_{\rv,l,\uparrow}c_{\rv+\yv^{(l)},l,\downarrow}+c_{\rv,l,\downarrow}c_{\rv+\yv^{(l)},l,\uparrow}
\end{aligned}
\end{equation}
where $\xv^{(l)}$ and $\yv^{(l)}$ are the lattice vectors on layer $l$.
The operators $\hat\Delta^{(1,2)}$ have $d_{x^2-y^2}$ symmetry within their own layer, in terms of the local axes.
However, they do not belong to irreducible representations of the $D_4$ point group of the bilayer.
That point group contains $\pi$-rotations with respect to axes lying in the plane, exchanging the two layers (see Fig.~2 or Ref.~\cite{lu_doping_2022}). 
Under these rotations, the operators $\hat\Delta^{(1,2)}$ are interchanged (with phases). 
We should therefore focus our attention on eigen-operators of the symmetry operations, which fall into the $B_1$ or the $B_2$ representation of the $D_4$ point group, defined as
\begin{equation}
\hat{B}_1=\hat{\Delta}^{(1)}+\hat{\Delta}^{(2)},\quad \hat{B}_2=\hat{\Delta}^{(1)}-\hat{\Delta}^{(2)}.
\end{equation}
These pairing operators are schematically illustrated on Fig.~\ref{fig:B1B2}. 
We use the convention from Ref.~\cite{lu_doping_2022}, where the links with the same sign in the $B_1$ representation are separated by the large angle (here corresponding to the complementary angle $90^\circ-\theta=46.4^\circ$). 
In the $B_2$ representation, the links with the same sign are separated by the smaller angle ($43.6^\circ$).  
At $45^\circ$, the two representations are equivalent and should correspond to degenerate states. 
The same is true if the inter-layer hopping $V$ vanishes.

%...............................................................................
%...............................................................................

A complex order parameter $\psi_a$ is defined as the average of the pairing operator per site
\begin{equation}
\psi_a=\frac{1}{L}\langle \hat{B}_a\rangle = \psi_a' + i\psi''_a \qquad (a=1,2)
\end{equation} 
where $\psi'_a$ and $\psi''_a$ are real and $L=58$ is the number of sites in the unit cell.
Both the $B_1$ and $B_2$ states may turn up as valid VCA solutions, but one of them will have a lower energy, depending on doping.
If only one of $\psi_{1,2}$ is nonzero, then its phase has no impact and can be set to zero ($\psi''_a=0$) without loss of generality.
However, in the TRS-breaking phase, the two states are close in energy, and the two order parameters may coexist in order to lower the energy further.
Since the $B_{1,2}$ states are symmetry eigenstates, the order parameters $\psi_{1,2}$ occur as scalars in the mean-field dispersion relation (the gap function) and the gap is evidently maximized when the two states are in quadrature, i.e., when the two order parameters have a $\pi/2$ phase difference. Since we have the freedom to set $\psi''_1=0$, this phase difference implies $\psi'_2=0$ and the full gap is then
$|\psi'_1+i\psi''_2| = \sqrt{(\psi'_1)^2+(\psi''_2)^2}$.
However, the phase difference $\phi$ between $\langle\hat{\Delta}^{(1)}\rangle$ and $\langle\hat{\Delta}^{(2)}\rangle$, i.e., the phase considered in the Josephson description of the bilayer, is different.
Since $\hat{\Delta}^{(1,2)} = (\hat B_1 \pm \hat B_2)/2$, then
\begin{align}
\phi & =\arg\frac{\psi_1-\psi_2}{\psi_1+\psi_2} 
= \arg\frac{\psi'_1-i\psi''_2}{\psi'_1+i\psi''_2} \notag\\
&= 2\arg(\psi'_1-i\psi''_2)
= -2\arctan\frac{\psi''_2}{\psi'_1}
\label{eq:phase}
\end{align}
A value of $\phi=0$ ($\phi=\pi$) corresponds to a pure $B_1$ ($B_2$) case. TRS breaking occurs when $\phi\neq 0$ or $\pi$.

To probe the TRS-breaking superconducting phase in the VCA, we use as Weiss fields the Hermitian operators $\hat B_1 + \hat B^\dagger_1$ and $i(\hat B^\dagger_2 - \hat B_2)$ simultaneously, and look for regions of doping where both are nonzero in the VCA solution. 
This corresponds then to the $B_1+iB_2$ state. 
The Weiss fields used in the VCA procedure are taken to be the same on all clusters (except on cluster $C$, where it is not defined). 
Using different Weiss fields on different clusters was tested; it did not improve the results significantly and dramatically increased the computational resources needed.  

Inter-layer pairing operators can also be defined. They were not used as Weiss fields (see Section IV of Ref.~\cite{lu_doping_2022} for an explanation). Nevertheless, inter-layer pairing occurs in the sense that the average value of the inter-layer pairing operators is nonzero.
This pairing propagates from the intra-layer pairing through inter-layer hopping.

%===============================================================================
\section{Result and discussion}
\label{sec:result58}
\subsection{TRS breaking phase}
In this section we consider the parameter describing Bi2212.
We start by probing the superconducting phase in model \eqref{eq:H} with $\theta=43.60^\circ$, for types $B_1$ and $B_2$ separately. 
We can compare the energies of the two states and find a doping region where TRS breaking is likely to occur. 

%...............................................................................
\begin{figure}[h]
	\begin{center}
	\includegraphics[width=\hsize]{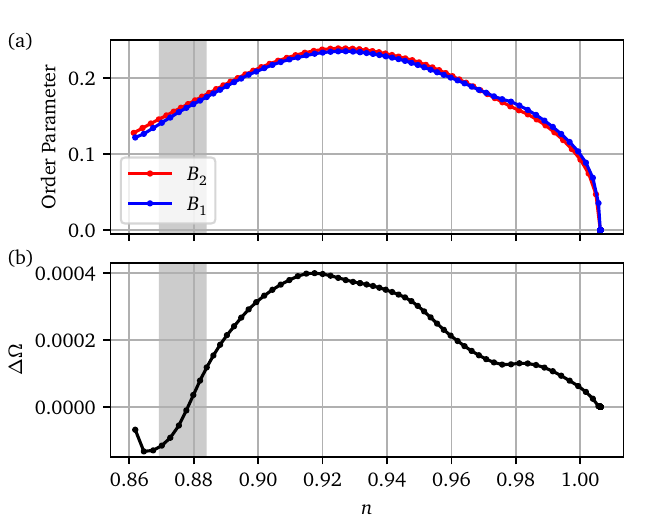}
	\caption{(a) : Norm of the order parameters $\psi_1$ and $\psi_2$ obtained from the VCA procedure with a Weiss field belonging respectively to the $B_1$ or $B_2$ representation, for an inter-layer hopping of  $V=0.4$, as a function of electron density $n$. (b) : Free energy difference $\Delta\Omega$ between the two states (in units of $t$). When $\Delta\Omega>0$ ($\Delta\Omega<0$) the $B_1$ ($B_2$) state is energetically favored. The grey region highlights the density range where a TRS breaking is observed (see text).
	}
	\label{fig:omega}
\end{center}
\end{figure}
%...............................................................................

On Fig.~\ref{fig:omega} (a), we show the superconducting order parameter for the $B_1$ and $B_2$ states for $V=0.4$. The two states have very similar superconducting domes. The order parameter vanishes around half-filling in both cases. That the order parameter does not vanish exactly at $n=1$ can be attributed to the error on the electron density typical of VCA when the chemical potential within the cluster is not treated as an additional variational parameter.

We can use the optimal value of the Potthoff functional in each state, $\Omega_{B_1}$ and $\Omega_{B_2}$, as a measure of the free energy of the system~\cite{potthoff_variational_2003}. In Fig.~\ref{fig:omega} (b), we show the difference $\Delta\Omega=\Omega_{B_2}-\Omega_{B_1}$. When $\Delta\Omega$ is positive (negative) the $B_1$ ($B_2$) representation is energetically favored. 

As shown in Fig.~\ref{fig:omega} (b), the difference $\Delta\Omega$ for $\theta=43.60^\circ$ is of order $10^{-4}$ (in unit of the nearest-neighbor hopping $t$). For $\theta=53.13^\circ$, the difference is more of order $10^{-3}$~\cite{lu_doping_2022}. This is expected since at $45^\circ$ the two representations should have the same energy. The closer the twist angle is to $45^\circ$, the closer in energy the states $B_1$ and $B_2$ should be and TRS breaking should be more likely. 
% In our numerical method, the energy level being this close makes it easier for a transition between both representations to happen. 
For $V=0.2$, we obtained similar results with an even smaller $\Delta\Omega$, owing to the smaller inter-layer tunneling.

\begin{figure}[h]
\begin{center}
\includegraphics[width=\hsize]{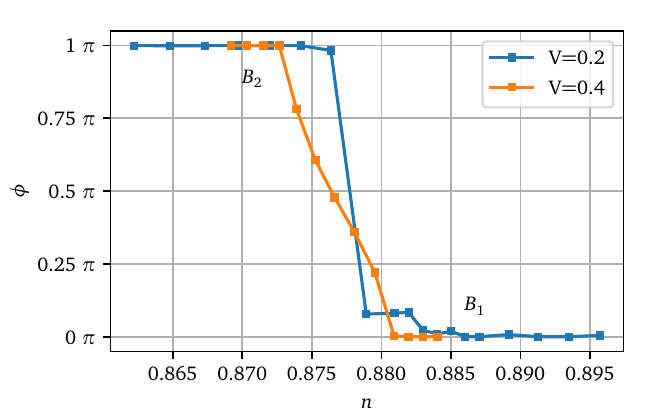}
\caption{Relative phase $\phi$ (Eq.~\eqref{eq:phase}) as a function of $n$. We can observe a transition from $B_1$ to $B_2$ when doping is increased. For $V=0.4$, a finite region of doping with $\phi$ between 0 and $\pi$ is observed, indicating a TRS-breaking phase.
For $V=0.2$, the width of this region, if it exists, is too small to be resolved.
}
\label{fig:phase_58sites}
\end{center}
\end{figure}

We then probed the superconducting phase when both the $B_1$ and $B_2$ states are simultaneously allowed. 
The two representations can now compete against each other or combine into a quadrature $B_1+iB_2$, whichever is energetically favorable. Fig.~\ref{fig:phase_58sites} shows the relative phase $\phi$ computed using Eq.~\eqref{eq:phase} as a function of the density $n$ in a narrow range of density. For $V=0.2$, it is possible to see an abrupt transition from the $B_1$ state to the $B_2$ state. The TRS-breaking region is too narrow to be seen with our method. The three data points with non-zero $\phi$ around $n=0.880$ may be an artifact caused by, e.g., the minimisation procedure, or the simplified Weiss field configuration used, with the same Weiss field on all clusters.

For $V=0.4$, the relative phase shows a continuous transition from $B_1$ to $B_2$ upon increasing hole doping. We obtain a TRS-breaking state between $n\approx0.873$ and $n\approx0.881$. The inter-layer tunneling amplitude is the most important parameter controlling the mixing of the two representations, as noted previously~\cite{lu_doping_2022}.

%...............................................................................
%...............................................................................

The grey area in Fig.~\ref{fig:omega} highlights the region where TRS breaking is seen in Fig.~\ref{fig:phase_58sites} for $V=0.4$.
It is well beyond optimal doping and corresponds to the range of doping where $\Delta\Omega$ is close to zero.

Our results are similar to those of Ref.~\cite{lu_doping_2022}, where a twist angle of $\theta=53.13^\circ$ was applied. 
A TRS-breaking phase near optimal doping was also observed in the strong inter-layer regime. 
The TRS-breaking phase observed with $\theta=43.60^\circ$ occurs within a smaller doping range, in the overdoped region. 
It was expected that, closer to $45^\circ$, the TRS-breaking phase would occur in a wider doping range. This is not what we observe here. 

To explain this, we turn to a simpler approach and consider, at $U=0$, the effect of a mean field associated with each of the superconducting states $B_1$ and $B_2$ for $\theta=43.60^\circ$. Usually we expect an energy gain when combining superconducting states from the same system if the nodes of each state are located at very different positions in the Brillouin zone (BZ). One example of this is the triplet state $p_x+ip_y$. This is not the case in twisted cuprates. One might think that combining a $d$-wave state from one layer with a $d$-wave from a second layer with a different orientation would effectively do the trick, but in fact this amounts to combining equivalent representations of two different systems instead of combining two different representations of the same system. In reality, the representations $B_1$ and $B_2$ have nodes at the same place.

\begin{figure}[h]
\begin{center}
\includegraphics[width=\hsize]{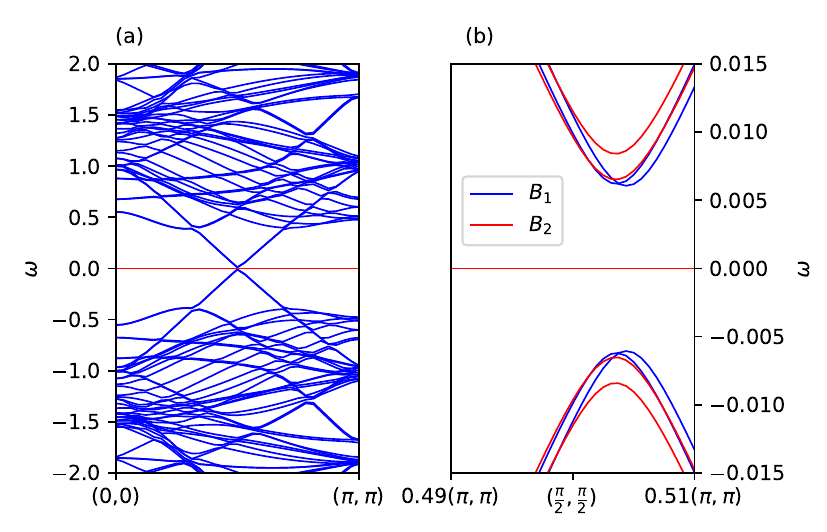}
\caption{ Band structure of model (\ref{eq:H}) with $U=0$ and a superconducting mean-field $b_i\hat{B}_i$. (a) Band structure along the diagonal of the first quadrant of the Brillouin zone for $b_1=0.2$. The nodes are located around $\mathbf{k}=(\pi/2,\pi/2)$. (b) Close-up of the region around the nodes for $b_1=0.2$ ( blue) and $b_2=0.2$ ( red). The nodes are in fact slightly gapped for both representations and very close to each other.
}
\label{fig:bands}
\end{center}
\end{figure}

In Fig \ref{fig:bands} (a) we show the band structure of Model (\ref{eq:H}) along the diagonal of the BZ for a non-zero $B_1$ mean-field ($b_1\hat{B}_1$, $b_1=0.2$). For a twist angle of $\theta=43.60^\circ$ there are 58 bands and the BZ is folded 29 times. There are nodes in the vicinity of $\mathbf{k}=(\pi/2,\pi/2)$ (same for the $B_2$ state, not shown in panel (a)). On panel (b), we compare the band structures with $B_1$ and $B_2$ mean-fields around the node (note the enlarged scale). The nodes are not perfect and are in fact already gapped for both states: no linear combination $B_1+iB_2$ is needed to generate the gap. Indeed, when a system has multiple orbitals, the nodes are not necessarily fixed by symmetry~\cite{kaba_group-theoretical_2019}. Following this argument, the energy gained by a complex combination $B_1+iB_2$ can remain small even close to $45^\circ$. 
Hence, the twist angle should not be the only parameter affecting the extent of the TRS-breaking phase, nor indeed the most important one. 
Other parameters, like doping (discussed in this work) and disorder~\cite{yuan_inhomogeneity-induced_2023}, play a key role.

%...............................................................................
%...............................................................................

\subsection{Effect of $U$ and $t'$}\label{sec:other_params}
\begin{figure}[h]
\begin{center}
\includegraphics[width=\hsize]{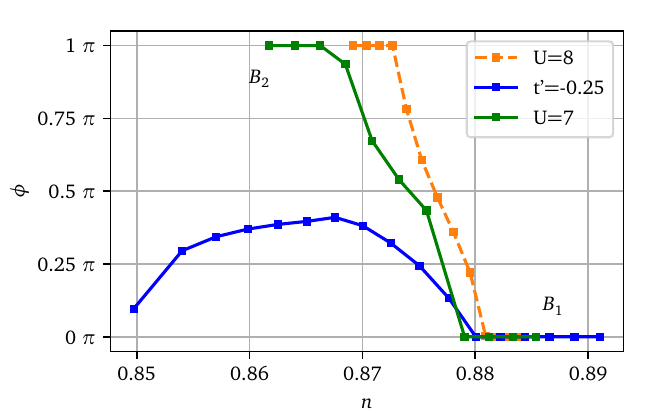}
\caption{Relative phase $\phi$ (Eq.~\eqref{eq:phase}) as a function of $n$ for three parameter sets. In all cases, $t=1$ and $V=0.4$. We can observe a finite region of doping with $\phi$ between 0 and $\pi$, indicating a TRS-breaking phase. For $t'=-0.25$, there is no transition between the two possible states: when hole doping is increased, the system starts from a $B_1$ state, then goes into a TRS-breaking phase, then back to the $B_1$ state; the doping range is increased when compared to the other data sets. For $U=0.7$, the behavior is closer to the $U=8$ case: there is a transition from the $B_1$ state to the $B_2$ as hole doping is increased. The data for $U=8$ are the same as those shown in Fig.~\ref{fig:phase_58sites}.\label{fig:phase_param} 
}

\end{center}
\end{figure}

The value of the second-neighbor hopping ($t'=-0.3$) used in the previous section was chosen to describe Bi2212 and to offer a comparison with previous work. We also applied our method to $t'=-0.25$ and $t'=-0.2$, other parameters staying the same ($t=1$, $U=8$, $V=0.4$).
The results for $t'=-0.25$ are shown on Fig.~\ref{fig:phase_param} (blue curve).
In that case the TRS-breaking phase also appears, but rather as an insertion within the $B_1$ phase instead of an intermediate state between $B_1$ and $B_2$.
The doping range over which it appears is also wider (about 3\%) but remains small compared to the one observed at $\theta=53.13^\circ$ in Ref.~\cite{lu_doping_2022}.
At $t'=-0.2$, the $B_1$ phase always has a lower energy than the $B_2$ phase and the TRS-breaking phase does not appear at all.
Hence a larger value of $|t'|$, which entails greater frustration of antiferromagnetic fluctuations, is beneficial to the TRS phase.
 
We also looked at the effect of the local repulsion $U$. 
The green curve on Fig.~\ref{fig:phase_param} shows the phase difference $\phi$ for $t=1$, $t'=-0.3$, $U=7$ and $V=0.4$ (same parameters as in the previous section, but with $U=7$ instead of $U=8$). In that case the TRS-breaking phase also occurs, but slightly shifted towards higher doping compared to $U=8$ (the original parameter set, shown in orange).
Additional computations at $U=10$ also revealed a TRS-breaking solution (not shown here).

Hence the existence of a TRS breaking phase is robust with respect to small changes in the model parameters.
However, the precise location of the TRS breaking phase is affected by the parameters of the single layer model, more so by the band parameters ($t'$) than by the interaction $U$.
In all cases the TRS phase is fragile: the associated energy gain at the twist angle considered here is of the order of $10^{-4}t$, which puts us in the sub-kelvin range.

%...............................................................................
%...............................................................................

% Fig. \ref{fig:phase_param} show the relative phase as a function of $n$ the electron density. It is possible to observe that the TRS breaking phase is present for all three sets of parameter. For a variation in $U$ the behavior of the relative phase is similar to the parameter set describing Bi2212. The system goes from a $B_1$ state to a $B_2$ state when doping is increased with a small region where the TRS is broken. We also looked at $U=10$ and found a TRS-breaking solution for some value of density (not shown here). For a variation in $t'$, the TRS breaking phase occur on a larger density range. Even though the range is bigger it is still relatively small, $\approx 3\%$ doping, compare to the result at $\theta=53.13^\circ$ from ref. \cite{lu_doping_2022}. For $t'=-0.25$, the system does not transition from a $B_1$ state to a $B_2$. This type of behavior was observed previously for a different system  \cite{lu_doping_2022}. All those, result indicate that the existence of a TRS breaking phase is robust to a small change in the parameter of the system.

%===============================================================================
\section{Conclusion}
\label{sec:conclusion}

We used a one-band Hubbard model for twisted bilayer cuprates at $\theta=43.60^\circ$ with VCA to search for a possible TRS-breaking superconducting phase. In the twisted bilayer, superconductivity is expected away from half filling either in the $B_1$ or $B_2$ representation of the $D_4$ point group. For a set of parameter chosen to describe Bi2212, we found that a phase transition from the $B_1$ state to the $B_2$ state when doping is increased. A region where the complex combination $B_1+iB_2$ is stable, thus breaking TRS, occurs in a small range of doping when the inter-layer doping is strong enough. Our results are in accord with previous work on the $53.13^\circ$ system, where the observation of TRS breaking needed a strong inter-layer tunneling~\cite{lu_doping_2022}. We also show that the two states $B_1$ and $B_2$ are closer in energy when the twist angle approaches $45^\circ$, where they are expected to be equivalent.

The energy difference between the two states being smaller as the twist angle approaches $45^\circ$ makes it harder to probe TRS breaking. With a simple mean-field argument, we show that the nodes of the $B_1$ and $B_2$ states are very close, making the energy gain from combining them very small. Hence, the twist angle cannot be the only parameter controlling the TRS breaking. The doping level needs to be considered, alongside the twist angle and probably disorder level, to create a robust TRS-breaking state.

We also looked at the effect of the model parameters. We find that the TRS breaking phase is robust against small variations of $t'$ and $U$, but that it is negatively impacted by smaller values of $|t'|$, i.e., by smaller AF frustration.
At the same time, the TRS phase only occurs on the overdoped side of the dome, i.e., on the more weakly correlated region of the phase diagram.
These two observations go in the same direction, i.e., strongly correlated superconductivity mediated by antiferromagnetic fluctuations is more hindrance than help to the TRS-breaking phase.
The TRS phase also requires a large interlayer hopping amplitude, larger than what can be expected in real materials, to be present in a significant doping range.

The system studied here is made of two layers of the same cuprate model, with the same doping. Taking inspiration from the heterobilayer transition metal dichalcogenides~\cite{ruiz-tijerina_interlayer_2019,tang_simulation_2020}, layers of different cuprates might be used to stabilize the TRS-breaking state. The difference in doping and structure might lead to a change in the TRS-breaking region. We will address this question in a future work.

%===============================================================================

\begin{acknowledgments}
This work was supported by the Natural Sciences and Engineering Research Council of Canada (NSERC) under grant RGPIN-2020-05060, by the NSERC postgraduate scholarships doctoral program and by the {\it Fonds de Recherche du Qu\'ebec Nature et technologies} (FRQNT) doctoral research scholarships.
Computational resources were provided by the Digital Research Alliance of Canada and Calcul Qu\'ebec.
\end{acknowledgments}

%===============================================================================
% \bibliography{twCuprate.bib}  % décommenter si biblio.bib existe

\begin{thebibliography}{42}%
\makeatletter
\providecommand \@ifxundefined [1]{%
 \@ifx{#1\undefined}
}%
\providecommand \@ifnum [1]{%
 \ifnum #1\expandafter \@firstoftwo
 \else \expandafter \@secondoftwo
 \fi
}%
\providecommand \@ifx [1]{%
 \ifx #1\expandafter \@firstoftwo
 \else \expandafter \@secondoftwo
 \fi
}%
\providecommand \natexlab [1]{#1}%
\providecommand \enquote  [1]{``#1''}%
\providecommand \bibnamefont  [1]{#1}%
\providecommand \bibfnamefont [1]{#1}%
\providecommand \citenamefont [1]{#1}%
\providecommand \href@noop [0]{\@secondoftwo}%
\providecommand \href [0]{\begingroup \@sanitize@url \@href}%
\providecommand \@href[1]{\@@startlink{#1}\@@href}%
\providecommand \@@href[1]{\endgroup#1\@@endlink}%
\providecommand \@sanitize@url [0]{\catcode `\\12\catcode `\$12\catcode `\&12\catcode `\#12\catcode `\^12\catcode `\_12\catcode `\%12\relax}%
\providecommand \@@startlink[1]{}%
\providecommand \@@endlink[0]{}%
\providecommand \url  [0]{\begingroup\@sanitize@url \@url }%
\providecommand \@url [1]{\endgroup\@href {#1}{\urlprefix }}%
\providecommand \urlprefix  [0]{URL }%
\providecommand \Eprint [0]{\href }%
\providecommand \doibase [0]{http://dx.doi.org/}%
\providecommand \selectlanguage [0]{\@gobble}%
\providecommand \bibinfo  [0]{\@secondoftwo}%
\providecommand \bibfield  [0]{\@secondoftwo}%
\providecommand \translation [1]{[#1]}%
\providecommand \BibitemOpen [0]{}%
\providecommand \bibitemStop [0]{}%
\providecommand \bibitemNoStop [0]{.\EOS\space}%
\providecommand \EOS [0]{\spacefactor3000\relax}%
\providecommand \BibitemShut  [1]{\csname bibitem#1\endcsname}%
\let\auto@bib@innerbib\@empty
%</preamble>
\bibitem [{\citenamefont {Cao}\ \emph {et~al.}(2018{\natexlab{a}})\citenamefont {Cao}, \citenamefont {Fatemi}, \citenamefont {Demir}, \citenamefont {Fang}, \citenamefont {Tomarken}, \citenamefont {Luo}, \citenamefont {Sanchez-Yamagishi}, \citenamefont {Watanabe}, \citenamefont {Taniguchi}, \citenamefont {Kaxiras}, \citenamefont {Ashoori},\ and\ \citenamefont {Jarillo-Herrero}}]{cao_correlated_2018}%
  \BibitemOpen
  \bibfield  {author} {\bibinfo {author} {\bibfnamefont {Y.}~\bibnamefont {Cao}}, \bibinfo {author} {\bibfnamefont {V.}~\bibnamefont {Fatemi}}, \bibinfo {author} {\bibfnamefont {A.}~\bibnamefont {Demir}}, \bibinfo {author} {\bibfnamefont {S.}~\bibnamefont {Fang}}, \bibinfo {author} {\bibfnamefont {S.~L.}\ \bibnamefont {Tomarken}}, \bibinfo {author} {\bibfnamefont {J.~Y.}\ \bibnamefont {Luo}}, \bibinfo {author} {\bibfnamefont {J.~D.}\ \bibnamefont {Sanchez-Yamagishi}}, \bibinfo {author} {\bibfnamefont {K.}~\bibnamefont {Watanabe}}, \bibinfo {author} {\bibfnamefont {T.}~\bibnamefont {Taniguchi}}, \bibinfo {author} {\bibfnamefont {E.}~\bibnamefont {Kaxiras}}, \bibinfo {author} {\bibfnamefont {R.~C.}\ \bibnamefont {Ashoori}}, \ and\ \bibinfo {author} {\bibfnamefont {P.}~\bibnamefont {Jarillo-Herrero}},\ }\href {\doibase 10.1038/nature26154} {\bibfield  {journal} {\bibinfo  {journal} {Nature}\ }\textbf {\bibinfo {volume} {556}},\ \bibinfo {pages} {80} (\bibinfo {year} {2018}{\natexlab{a}})}\BibitemShut {NoStop}%
\bibitem [{\citenamefont {Cao}\ \emph {et~al.}(2018{\natexlab{b}})\citenamefont {Cao}, \citenamefont {Fatemi}, \citenamefont {Fang}, \citenamefont {Watanabe}, \citenamefont {Taniguchi}, \citenamefont {Kaxiras},\ and\ \citenamefont {Jarillo-Herrero}}]{cao_unconventional_2018}%
  \BibitemOpen
  \bibfield  {author} {\bibinfo {author} {\bibfnamefont {Y.}~\bibnamefont {Cao}}, \bibinfo {author} {\bibfnamefont {V.}~\bibnamefont {Fatemi}}, \bibinfo {author} {\bibfnamefont {S.}~\bibnamefont {Fang}}, \bibinfo {author} {\bibfnamefont {K.}~\bibnamefont {Watanabe}}, \bibinfo {author} {\bibfnamefont {T.}~\bibnamefont {Taniguchi}}, \bibinfo {author} {\bibfnamefont {E.}~\bibnamefont {Kaxiras}}, \ and\ \bibinfo {author} {\bibfnamefont {P.}~\bibnamefont {Jarillo-Herrero}},\ }\href {\doibase 10.1038/nature26160} {\bibfield  {journal} {\bibinfo  {journal} {Nature}\ }\textbf {\bibinfo {volume} {556}},\ \bibinfo {pages} {43} (\bibinfo {year} {2018}{\natexlab{b}})}\BibitemShut {NoStop}%
\bibitem [{\citenamefont {Balents}\ \emph {et~al.}(2020)\citenamefont {Balents}, \citenamefont {Dean}, \citenamefont {Efetov},\ and\ \citenamefont {Young}}]{balents_superconductivity_2020}%
  \BibitemOpen
  \bibfield  {author} {\bibinfo {author} {\bibfnamefont {L.}~\bibnamefont {Balents}}, \bibinfo {author} {\bibfnamefont {C.~R.}\ \bibnamefont {Dean}}, \bibinfo {author} {\bibfnamefont {D.~K.}\ \bibnamefont {Efetov}}, \ and\ \bibinfo {author} {\bibfnamefont {A.~F.}\ \bibnamefont {Young}},\ }\href {\doibase 10.1038/s41567-020-0906-9} {\bibfield  {journal} {\bibinfo  {journal} {Nat. Phys.}\ }\textbf {\bibinfo {volume} {16}},\ \bibinfo {pages} {725} (\bibinfo {year} {2020})}\BibitemShut {NoStop}%
\bibitem [{\citenamefont {Xian}\ \emph {et~al.}(2019)\citenamefont {Xian}, \citenamefont {Kennes}, \citenamefont {Tancogne-Dejean}, \citenamefont {Altarelli},\ and\ \citenamefont {Rubio}}]{xian_multiflat_2019}%
  \BibitemOpen
  \bibfield  {author} {\bibinfo {author} {\bibfnamefont {L.}~\bibnamefont {Xian}}, \bibinfo {author} {\bibfnamefont {D.~M.}\ \bibnamefont {Kennes}}, \bibinfo {author} {\bibfnamefont {N.}~\bibnamefont {Tancogne-Dejean}}, \bibinfo {author} {\bibfnamefont {M.}~\bibnamefont {Altarelli}}, \ and\ \bibinfo {author} {\bibfnamefont {A.}~\bibnamefont {Rubio}},\ }\href {\doibase 10.1021/acs.nanolett.9b00986} {\bibfield  {journal} {\bibinfo  {journal} {Nano Lett.}\ }\textbf {\bibinfo {volume} {19}},\ \bibinfo {pages} {4934} (\bibinfo {year} {2019})}\BibitemShut {NoStop}%
\bibitem [{\citenamefont {Ruiz-Tijerina}\ and\ \citenamefont {Fal'ko}(2019)}]{ruiz-tijerina_interlayer_2019}%
  \BibitemOpen
  \bibfield  {author} {\bibinfo {author} {\bibfnamefont {D.~A.}\ \bibnamefont {Ruiz-Tijerina}}\ and\ \bibinfo {author} {\bibfnamefont {V.~I.}\ \bibnamefont {Fal'ko}},\ }\href {\doibase 10.1103/PhysRevB.99.125424} {\bibfield  {journal} {\bibinfo  {journal} {Phys. Rev. B}\ }\textbf {\bibinfo {volume} {99}},\ \bibinfo {pages} {125424} (\bibinfo {year} {2019})}\BibitemShut {NoStop}%
\bibitem [{\citenamefont {Wang}\ \emph {et~al.}(2020)\citenamefont {Wang}, \citenamefont {Shih}, \citenamefont {Ghiotto}, \citenamefont {Xian}, \citenamefont {Rhodes}, \citenamefont {Tan}, \citenamefont {Claassen}, \citenamefont {Kennes}, \citenamefont {Bai}, \citenamefont {Kim}, \citenamefont {Watanabe}, \citenamefont {Taniguchi}, \citenamefont {Zhu}, \citenamefont {Hone}, \citenamefont {Rubio}, \citenamefont {Pasupathy},\ and\ \citenamefont {Dean}}]{wang_correlated_2020}%
  \BibitemOpen
  \bibfield  {author} {\bibinfo {author} {\bibfnamefont {L.}~\bibnamefont {Wang}}, \bibinfo {author} {\bibfnamefont {E.-M.}\ \bibnamefont {Shih}}, \bibinfo {author} {\bibfnamefont {A.}~\bibnamefont {Ghiotto}}, \bibinfo {author} {\bibfnamefont {L.}~\bibnamefont {Xian}}, \bibinfo {author} {\bibfnamefont {D.~A.}\ \bibnamefont {Rhodes}}, \bibinfo {author} {\bibfnamefont {C.}~\bibnamefont {Tan}}, \bibinfo {author} {\bibfnamefont {M.}~\bibnamefont {Claassen}}, \bibinfo {author} {\bibfnamefont {D.~M.}\ \bibnamefont {Kennes}}, \bibinfo {author} {\bibfnamefont {Y.}~\bibnamefont {Bai}}, \bibinfo {author} {\bibfnamefont {B.}~\bibnamefont {Kim}}, \bibinfo {author} {\bibfnamefont {K.}~\bibnamefont {Watanabe}}, \bibinfo {author} {\bibfnamefont {T.}~\bibnamefont {Taniguchi}}, \bibinfo {author} {\bibfnamefont {X.}~\bibnamefont {Zhu}}, \bibinfo {author} {\bibfnamefont {J.}~\bibnamefont {Hone}}, \bibinfo {author} {\bibfnamefont {A.}~\bibnamefont {Rubio}}, \bibinfo {author} {\bibfnamefont {A.~N.}\ \bibnamefont {Pasupathy}}, \ and\ \bibinfo {author} {\bibfnamefont {C.~R.}\ \bibnamefont {Dean}},\ }\href {\doibase 10.1038/s41563-020-0708-6} {\bibfield  {journal} {\bibinfo  {journal} {Nat. Mater.}\ }\textbf {\bibinfo {volume} {19}},\ \bibinfo {pages} {861} (\bibinfo {year} {2020})}\BibitemShut {NoStop}%
\bibitem [{\citenamefont {An}\ \emph {et~al.}(2020)\citenamefont {An}, \citenamefont {Cai}, \citenamefont {Pei}, \citenamefont {Huang}, \citenamefont {Wu}, \citenamefont {Zhou}, \citenamefont {Lin}, \citenamefont {Ying}, \citenamefont {Ye}, \citenamefont {Feng}, \citenamefont {Gao}, \citenamefont {Cacho}, \citenamefont {Watson}, \citenamefont {Chen},\ and\ \citenamefont {Wang}}]{an_interaction_2020}%
  \BibitemOpen
  \bibfield  {author} {\bibinfo {author} {\bibfnamefont {L.}~\bibnamefont {An}}, \bibinfo {author} {\bibfnamefont {X.}~\bibnamefont {Cai}}, \bibinfo {author} {\bibfnamefont {D.}~\bibnamefont {Pei}}, \bibinfo {author} {\bibfnamefont {M.}~\bibnamefont {Huang}}, \bibinfo {author} {\bibfnamefont {Z.}~\bibnamefont {Wu}}, \bibinfo {author} {\bibfnamefont {Z.}~\bibnamefont {Zhou}}, \bibinfo {author} {\bibfnamefont {J.}~\bibnamefont {Lin}}, \bibinfo {author} {\bibfnamefont {Z.}~\bibnamefont {Ying}}, \bibinfo {author} {\bibfnamefont {Z.}~\bibnamefont {Ye}}, \bibinfo {author} {\bibfnamefont {X.}~\bibnamefont {Feng}}, \bibinfo {author} {\bibfnamefont {R.}~\bibnamefont {Gao}}, \bibinfo {author} {\bibfnamefont {C.}~\bibnamefont {Cacho}}, \bibinfo {author} {\bibfnamefont {M.}~\bibnamefont {Watson}}, \bibinfo {author} {\bibfnamefont {Y.}~\bibnamefont {Chen}}, \ and\ \bibinfo {author} {\bibfnamefont {N.}~\bibnamefont {Wang}},\ }\href {\doibase 10.1039/D0NH00248H} {\bibfield  {journal} {\bibinfo  {journal} {Nanoscale Horiz.}\ }\textbf {\bibinfo {volume} {5}},\ \bibinfo {pages} {1309} (\bibinfo {year} {2020})}\BibitemShut {NoStop}%
\bibitem [{\citenamefont {Zhang}\ \emph {et~al.}(2020)\citenamefont {Zhang}, \citenamefont {Wang}, \citenamefont {Watanabe}, \citenamefont {Taniguchi}, \citenamefont {Ueno}, \citenamefont {Tutuc},\ and\ \citenamefont {LeRoy}}]{zhang_flat_2020}%
  \BibitemOpen
  \bibfield  {author} {\bibinfo {author} {\bibfnamefont {Z.}~\bibnamefont {Zhang}}, \bibinfo {author} {\bibfnamefont {Y.}~\bibnamefont {Wang}}, \bibinfo {author} {\bibfnamefont {K.}~\bibnamefont {Watanabe}}, \bibinfo {author} {\bibfnamefont {T.}~\bibnamefont {Taniguchi}}, \bibinfo {author} {\bibfnamefont {K.}~\bibnamefont {Ueno}}, \bibinfo {author} {\bibfnamefont {E.}~\bibnamefont {Tutuc}}, \ and\ \bibinfo {author} {\bibfnamefont {B.~J.}\ \bibnamefont {LeRoy}},\ }\href {\doibase 10.1038/s41567-020-0958-x} {\bibfield  {journal} {\bibinfo  {journal} {Nat. Phys.}\ }\textbf {\bibinfo {volume} {16}},\ \bibinfo {pages} {1093} (\bibinfo {year} {2020})}\BibitemShut {NoStop}%
\bibitem [{\citenamefont {Venkateswarlu}\ \emph {et~al.}(2020)\citenamefont {Venkateswarlu}, \citenamefont {Honecker},\ and\ \citenamefont {Trambly~de Laissardière}}]{venkateswarlu_electronic_2020}%
  \BibitemOpen
  \bibfield  {author} {\bibinfo {author} {\bibfnamefont {S.}~\bibnamefont {Venkateswarlu}}, \bibinfo {author} {\bibfnamefont {A.}~\bibnamefont {Honecker}}, \ and\ \bibinfo {author} {\bibfnamefont {G.}~\bibnamefont {Trambly~de Laissardière}},\ }\href {\doibase 10.1103/PhysRevB.102.081103} {\bibfield  {journal} {\bibinfo  {journal} {Phys. Rev. B}\ }\textbf {\bibinfo {volume} {102}},\ \bibinfo {pages} {081103} (\bibinfo {year} {2020})}\BibitemShut {NoStop}%
\bibitem [{\citenamefont {Regan}\ \emph {et~al.}(2020)\citenamefont {Regan}, \citenamefont {Wang}, \citenamefont {Jin}, \citenamefont {Bakti~Utama}, \citenamefont {Gao}, \citenamefont {Wei}, \citenamefont {Zhao}, \citenamefont {Zhao}, \citenamefont {Zhang}, \citenamefont {Yumigeta}, \citenamefont {Blei}, \citenamefont {Carlström}, \citenamefont {Watanabe}, \citenamefont {Taniguchi}, \citenamefont {Tongay}, \citenamefont {Crommie}, \citenamefont {Zettl},\ and\ \citenamefont {Wang}}]{regan_mott_2020}%
  \BibitemOpen
  \bibfield  {author} {\bibinfo {author} {\bibfnamefont {E.~C.}\ \bibnamefont {Regan}}, \bibinfo {author} {\bibfnamefont {D.}~\bibnamefont {Wang}}, \bibinfo {author} {\bibfnamefont {C.}~\bibnamefont {Jin}}, \bibinfo {author} {\bibfnamefont {M.~I.}\ \bibnamefont {Bakti~Utama}}, \bibinfo {author} {\bibfnamefont {B.}~\bibnamefont {Gao}}, \bibinfo {author} {\bibfnamefont {X.}~\bibnamefont {Wei}}, \bibinfo {author} {\bibfnamefont {S.}~\bibnamefont {Zhao}}, \bibinfo {author} {\bibfnamefont {W.}~\bibnamefont {Zhao}}, \bibinfo {author} {\bibfnamefont {Z.}~\bibnamefont {Zhang}}, \bibinfo {author} {\bibfnamefont {K.}~\bibnamefont {Yumigeta}}, \bibinfo {author} {\bibfnamefont {M.}~\bibnamefont {Blei}}, \bibinfo {author} {\bibfnamefont {J.~D.}\ \bibnamefont {Carlström}}, \bibinfo {author} {\bibfnamefont {K.}~\bibnamefont {Watanabe}}, \bibinfo {author} {\bibfnamefont {T.}~\bibnamefont {Taniguchi}}, \bibinfo {author} {\bibfnamefont {S.}~\bibnamefont {Tongay}}, \bibinfo {author} {\bibfnamefont {M.}~\bibnamefont {Crommie}}, \bibinfo {author} {\bibfnamefont {A.}~\bibnamefont {Zettl}}, \ and\ \bibinfo {author} {\bibfnamefont {F.}~\bibnamefont {Wang}},\ }\href {\doibase 10.1038/s41586-020-2092-4} {\bibfield  {journal} {\bibinfo  {journal} {Nature}\ }\textbf {\bibinfo {volume} {579}},\ \bibinfo {pages} {359} (\bibinfo {year} {2020})}\BibitemShut {NoStop}%
\bibitem [{\citenamefont {Tang}\ \emph {et~al.}(2020)\citenamefont {Tang}, \citenamefont {Li}, \citenamefont {Li}, \citenamefont {Xu}, \citenamefont {Liu}, \citenamefont {Barmak}, \citenamefont {Watanabe}, \citenamefont {Taniguchi}, \citenamefont {MacDonald}, \citenamefont {Shan},\ and\ \citenamefont {Mak}}]{tang_simulation_2020}%
  \BibitemOpen
  \bibfield  {author} {\bibinfo {author} {\bibfnamefont {Y.}~\bibnamefont {Tang}}, \bibinfo {author} {\bibfnamefont {L.}~\bibnamefont {Li}}, \bibinfo {author} {\bibfnamefont {T.}~\bibnamefont {Li}}, \bibinfo {author} {\bibfnamefont {Y.}~\bibnamefont {Xu}}, \bibinfo {author} {\bibfnamefont {S.}~\bibnamefont {Liu}}, \bibinfo {author} {\bibfnamefont {K.}~\bibnamefont {Barmak}}, \bibinfo {author} {\bibfnamefont {K.}~\bibnamefont {Watanabe}}, \bibinfo {author} {\bibfnamefont {T.}~\bibnamefont {Taniguchi}}, \bibinfo {author} {\bibfnamefont {A.~H.}\ \bibnamefont {MacDonald}}, \bibinfo {author} {\bibfnamefont {J.}~\bibnamefont {Shan}}, \ and\ \bibinfo {author} {\bibfnamefont {K.~F.}\ \bibnamefont {Mak}},\ }\href {\doibase 10.1038/s41586-020-2085-3} {\bibfield  {journal} {\bibinfo  {journal} {Nature}\ }\textbf {\bibinfo {volume} {579}},\ \bibinfo {pages} {353} (\bibinfo {year} {2020})}\BibitemShut {NoStop}%
\bibitem [{\citenamefont {Bélanger}\ \emph {et~al.}(2022)\citenamefont {Bélanger}, \citenamefont {Fournier},\ and\ \citenamefont {Sénéchal}}]{belanger_superconductivity_2022}%
  \BibitemOpen
  \bibfield  {author} {\bibinfo {author} {\bibfnamefont {M.}~\bibnamefont {Bélanger}}, \bibinfo {author} {\bibfnamefont {J.}~\bibnamefont {Fournier}}, \ and\ \bibinfo {author} {\bibfnamefont {D.}~\bibnamefont {Sénéchal}},\ }\href {\doibase 10.1103/PhysRevB.106.235135} {\bibfield  {journal} {\bibinfo  {journal} {Phys. Rev. B}\ }\textbf {\bibinfo {volume} {106}},\ \bibinfo {pages} {235135} (\bibinfo {year} {2022})}\BibitemShut {NoStop}%
\bibitem [{\citenamefont {Yu}\ \emph {et~al.}(2019)\citenamefont {Yu}, \citenamefont {Ma}, \citenamefont {Cai}, \citenamefont {Zhong}, \citenamefont {Ye}, \citenamefont {Shen}, \citenamefont {Gu}, \citenamefont {Chen},\ and\ \citenamefont {Zhang}}]{yu_high-temperature_2019}%
  \BibitemOpen
  \bibfield  {author} {\bibinfo {author} {\bibfnamefont {Y.}~\bibnamefont {Yu}}, \bibinfo {author} {\bibfnamefont {L.}~\bibnamefont {Ma}}, \bibinfo {author} {\bibfnamefont {P.}~\bibnamefont {Cai}}, \bibinfo {author} {\bibfnamefont {R.}~\bibnamefont {Zhong}}, \bibinfo {author} {\bibfnamefont {C.}~\bibnamefont {Ye}}, \bibinfo {author} {\bibfnamefont {J.}~\bibnamefont {Shen}}, \bibinfo {author} {\bibfnamefont {G.~D.}\ \bibnamefont {Gu}}, \bibinfo {author} {\bibfnamefont {X.~H.}\ \bibnamefont {Chen}}, \ and\ \bibinfo {author} {\bibfnamefont {Y.}~\bibnamefont {Zhang}},\ }\href {\doibase 10.1038/s41586-019-1718-x} {\bibfield  {journal} {\bibinfo  {journal} {Nature}\ }\textbf {\bibinfo {volume} {575}},\ \bibinfo {pages} {156} (\bibinfo {year} {2019})}\BibitemShut {NoStop}%
\bibitem [{\citenamefont {Zhao}\ \emph {et~al.}(2019)\citenamefont {Zhao}, \citenamefont {Poccia}, \citenamefont {Panetta}, \citenamefont {Yu}, \citenamefont {Johnson}, \citenamefont {Yoo}, \citenamefont {Zhong}, \citenamefont {Gu}, \citenamefont {Watanabe}, \citenamefont {Taniguchi}, \citenamefont {Postolova}, \citenamefont {Vinokur},\ and\ \citenamefont {Kim}}]{zhao_sign-reversing_2019}%
  \BibitemOpen
  \bibfield  {author} {\bibinfo {author} {\bibfnamefont {S.~F.}\ \bibnamefont {Zhao}}, \bibinfo {author} {\bibfnamefont {N.}~\bibnamefont {Poccia}}, \bibinfo {author} {\bibfnamefont {M.~G.}\ \bibnamefont {Panetta}}, \bibinfo {author} {\bibfnamefont {C.}~\bibnamefont {Yu}}, \bibinfo {author} {\bibfnamefont {J.~W.}\ \bibnamefont {Johnson}}, \bibinfo {author} {\bibfnamefont {H.}~\bibnamefont {Yoo}}, \bibinfo {author} {\bibfnamefont {R.}~\bibnamefont {Zhong}}, \bibinfo {author} {\bibfnamefont {G.}~\bibnamefont {Gu}}, \bibinfo {author} {\bibfnamefont {K.}~\bibnamefont {Watanabe}}, \bibinfo {author} {\bibfnamefont {T.}~\bibnamefont {Taniguchi}}, \bibinfo {author} {\bibfnamefont {S.~V.}\ \bibnamefont {Postolova}}, \bibinfo {author} {\bibfnamefont {V.~M.}\ \bibnamefont {Vinokur}}, \ and\ \bibinfo {author} {\bibfnamefont {P.}~\bibnamefont {Kim}},\ }\href {\doibase 10.1103/PhysRevLett.122.247001} {\bibfield  {journal} {\bibinfo  {journal} {Phys. Rev. Lett.}\ }\textbf {\bibinfo {volume} {122}},\ \bibinfo {pages} {247001} (\bibinfo {year} {2019})}\BibitemShut {NoStop}%
\bibitem [{\citenamefont {Can}\ \emph {et~al.}(2021)\citenamefont {Can}, \citenamefont {Tummuru}, \citenamefont {Day}, \citenamefont {Elfimov}, \citenamefont {Damascelli},\ and\ \citenamefont {Franz}}]{can_high-temperature_2021}%
  \BibitemOpen
  \bibfield  {author} {\bibinfo {author} {\bibfnamefont {O.}~\bibnamefont {Can}}, \bibinfo {author} {\bibfnamefont {T.}~\bibnamefont {Tummuru}}, \bibinfo {author} {\bibfnamefont {R.~P.}\ \bibnamefont {Day}}, \bibinfo {author} {\bibfnamefont {I.}~\bibnamefont {Elfimov}}, \bibinfo {author} {\bibfnamefont {A.}~\bibnamefont {Damascelli}}, \ and\ \bibinfo {author} {\bibfnamefont {M.}~\bibnamefont {Franz}},\ }\href {\doibase 10.1038/s41567-020-01142-7} {\bibfield  {journal} {\bibinfo  {journal} {Nat. Phys.}\ }\textbf {\bibinfo {volume} {17}},\ \bibinfo {pages} {519} (\bibinfo {year} {2021})}\BibitemShut {NoStop}%
\bibitem [{\citenamefont {Volkov}\ \emph {et~al.}(2023)\citenamefont {Volkov}, \citenamefont {Wilson}, \citenamefont {Lucht},\ and\ \citenamefont {Pixley}}]{volkov_magic_2023}%
  \BibitemOpen
  \bibfield  {author} {\bibinfo {author} {\bibfnamefont {P.~A.}\ \bibnamefont {Volkov}}, \bibinfo {author} {\bibfnamefont {J.~H.}\ \bibnamefont {Wilson}}, \bibinfo {author} {\bibfnamefont {K.~P.}\ \bibnamefont {Lucht}}, \ and\ \bibinfo {author} {\bibfnamefont {J.~H.}\ \bibnamefont {Pixley}},\ }\href {\doibase 10.1103/PhysRevB.107.174506} {\bibfield  {journal} {\bibinfo  {journal} {Phys. Rev. B}\ }\textbf {\bibinfo {volume} {107}},\ \bibinfo {pages} {174506} (\bibinfo {year} {2023})}\BibitemShut {NoStop}%
\bibitem [{\citenamefont {Lu}\ and\ \citenamefont {Sénéchal}(2022)}]{lu_doping_2022}%
  \BibitemOpen
  \bibfield  {author} {\bibinfo {author} {\bibfnamefont {X.}~\bibnamefont {Lu}}\ and\ \bibinfo {author} {\bibfnamefont {D.}~\bibnamefont {Sénéchal}},\ }\href {\doibase 10.1103/PhysRevB.105.245127} {\bibfield  {journal} {\bibinfo  {journal} {Phys. Rev. B}\ }\textbf {\bibinfo {volume} {105}},\ \bibinfo {pages} {245127} (\bibinfo {year} {2022})}\BibitemShut {NoStop}%
\bibitem [{\citenamefont {Margalit}\ \emph {et~al.}(2022)\citenamefont {Margalit}, \citenamefont {Yan}, \citenamefont {Franz},\ and\ \citenamefont {Oreg}}]{margalit_chiral_2022}%
  \BibitemOpen
  \bibfield  {author} {\bibinfo {author} {\bibfnamefont {G.}~\bibnamefont {Margalit}}, \bibinfo {author} {\bibfnamefont {B.}~\bibnamefont {Yan}}, \bibinfo {author} {\bibfnamefont {M.}~\bibnamefont {Franz}}, \ and\ \bibinfo {author} {\bibfnamefont {Y.}~\bibnamefont {Oreg}},\ }\href {\doibase 10.1103/PhysRevB.106.205424} {\bibfield  {journal} {\bibinfo  {journal} {Phys. Rev. B}\ }\textbf {\bibinfo {volume} {106}},\ \bibinfo {pages} {205424} (\bibinfo {year} {2022})},\ \bibinfo {note} {arXiv:2209.11080 [cond-mat]}\BibitemShut {NoStop}%
\bibitem [{\citenamefont {Mercado}\ \emph {et~al.}(2022)\citenamefont {Mercado}, \citenamefont {Sahoo},\ and\ \citenamefont {Franz}}]{mercado_high-temperature_2022}%
  \BibitemOpen
  \bibfield  {author} {\bibinfo {author} {\bibfnamefont {A.}~\bibnamefont {Mercado}}, \bibinfo {author} {\bibfnamefont {S.}~\bibnamefont {Sahoo}}, \ and\ \bibinfo {author} {\bibfnamefont {M.}~\bibnamefont {Franz}},\ }\href {\doibase 10.1103/PhysRevLett.128.137002} {\bibfield  {journal} {\bibinfo  {journal} {Phys. Rev. Lett.}\ }\textbf {\bibinfo {volume} {128}},\ \bibinfo {pages} {137002} (\bibinfo {year} {2022})}\BibitemShut {NoStop}%
\bibitem [{\citenamefont {Tummuru}\ \emph {et~al.}(2022{\natexlab{a}})\citenamefont {Tummuru}, \citenamefont {Lantagne-Hurtubise},\ and\ \citenamefont {Franz}}]{tummuru_twisted_2022}%
  \BibitemOpen
  \bibfield  {author} {\bibinfo {author} {\bibfnamefont {T.}~\bibnamefont {Tummuru}}, \bibinfo {author} {\bibfnamefont {\'E.}~\bibnamefont {Lantagne-Hurtubise}}, \ and\ \bibinfo {author} {\bibfnamefont {M.}~\bibnamefont {Franz}},\ }\href {\doibase 10.1103/PhysRevB.106.014520} {\bibfield  {journal} {\bibinfo  {journal} {Phys. Rev. B}\ }\textbf {\bibinfo {volume} {106}},\ \bibinfo {pages} {014520} (\bibinfo {year} {2022}{\natexlab{a}})}\BibitemShut {NoStop}%
\bibitem [{\citenamefont {Martini}\ \emph {et~al.}(2023)\citenamefont
  {Martini}, \citenamefont {Lee}, \citenamefont {Confalone}, \citenamefont
  {Shokri}, \citenamefont {Saggau}, \citenamefont {Wolf}, \citenamefont {Gu},
  \citenamefont {Watanabe}, \citenamefont {Taniguchi}, \citenamefont
  {Montemurro}, \citenamefont {Vinokur}, \citenamefont {Nielsch},\ and\
  \citenamefont {Poccia}}]{martini_twisted_2023}%
  \BibitemOpen
  \bibfield  {author} {\bibinfo {author} {\bibfnamefont {M.}~\bibnamefont
  {Martini}}, \bibinfo {author} {\bibfnamefont {Y.}~\bibnamefont {Lee}},
  \bibinfo {author} {\bibfnamefont {T.}~\bibnamefont {Confalone}}, \bibinfo
  {author} {\bibfnamefont {S.}~\bibnamefont {Shokri}}, \bibinfo {author}
  {\bibfnamefont {C.~N.}\ \bibnamefont {Saggau}}, \bibinfo {author}
  {\bibfnamefont {D.}~\bibnamefont {Wolf}}, \bibinfo {author} {\bibfnamefont
  {G.}~\bibnamefont {Gu}}, \bibinfo {author} {\bibfnamefont {K.}~\bibnamefont
  {Watanabe}}, \bibinfo {author} {\bibfnamefont {T.}~\bibnamefont {Taniguchi}},
  \bibinfo {author} {\bibfnamefont {D.}~\bibnamefont {Montemurro}}, \bibinfo
  {author} {\bibfnamefont {V.~M.}\ \bibnamefont {Vinokur}}, \bibinfo {author}
  {\bibfnamefont {K.}~\bibnamefont {Nielsch}}, \ and\ \bibinfo {author}
  {\bibfnamefont {N.}~\bibnamefont {Poccia}},\ }\href {\doibase
  10.1016/j.mattod.2023.06.007} {\bibfield  {journal} {\bibinfo  {journal}
  {Materials Today}\ }\textbf {\bibinfo {volume} {67}},\ \bibinfo {pages} {106}
  (\bibinfo {year} {2023})}\BibitemShut {NoStop}%
\bibitem [{\citenamefont {Lee}\ \emph {et~al.}(2023)\citenamefont {Lee}, \citenamefont {Martini}, \citenamefont {Confalone}, \citenamefont {Shokri}, \citenamefont {Saggau}, \citenamefont {Wolf}, \citenamefont {Gu}, \citenamefont {Watanabe}, \citenamefont {Taniguchi}, \citenamefont {Montemurro}, \citenamefont {Vinokur}, \citenamefont {Nielsch},\ and\ \citenamefont {Poccia}}]{lee_encapsulating_2023}%
  \BibitemOpen
  \bibfield  {author} {\bibinfo {author} {\bibfnamefont {Y.}~\bibnamefont {Lee}}, \bibinfo {author} {\bibfnamefont {M.}~\bibnamefont {Martini}}, \bibinfo {author} {\bibfnamefont {T.}~\bibnamefont {Confalone}}, \bibinfo {author} {\bibfnamefont {S.}~\bibnamefont {Shokri}}, \bibinfo {author} {\bibfnamefont {C.~N.}\ \bibnamefont {Saggau}}, \bibinfo {author} {\bibfnamefont {D.}~\bibnamefont {Wolf}}, \bibinfo {author} {\bibfnamefont {G.}~\bibnamefont {Gu}}, \bibinfo {author} {\bibfnamefont {K.}~\bibnamefont {Watanabe}}, \bibinfo {author} {\bibfnamefont {T.}~\bibnamefont {Taniguchi}}, \bibinfo {author} {\bibfnamefont {D.}~\bibnamefont {Montemurro}}, \bibinfo {author} {\bibfnamefont {V.~M.}\ \bibnamefont {Vinokur}}, \bibinfo {author} {\bibfnamefont {K.}~\bibnamefont {Nielsch}}, \ and\ \bibinfo {author} {\bibfnamefont {N.}~\bibnamefont {Poccia}},\ }\href {\doibase https://doi.org/10.1002/adma.202209135} {\bibfield  {journal} {\bibinfo  {journal} {Advanced Materials}\ }\textbf {\bibinfo {volume} {35}},\ \bibinfo {pages} {2209135} (\bibinfo {year} {2023})}\BibitemShut {NoStop}%
\bibitem [{\citenamefont {Zhu}\ \emph {et~al.}(2021)\citenamefont {Zhu}, \citenamefont {Liao}, \citenamefont {Zhang}, \citenamefont {Xie}, \citenamefont {Meng}, \citenamefont {Liu}, \citenamefont {Bai}, \citenamefont {Ji}, \citenamefont {Zhang}, \citenamefont {Jiang}, \citenamefont {Zhong}, \citenamefont {Schneeloch}, \citenamefont {Gu}, \citenamefont {Gu}, \citenamefont {Ma}, \citenamefont {Zhang},\ and\ \citenamefont {Xue}}]{zhu_presence_2021}%
  \BibitemOpen
  \bibfield  {author} {\bibinfo {author} {\bibfnamefont {Y.}~\bibnamefont {Zhu}}, \bibinfo {author} {\bibfnamefont {M.}~\bibnamefont {Liao}}, \bibinfo {author} {\bibfnamefont {Q.}~\bibnamefont {Zhang}}, \bibinfo {author} {\bibfnamefont {H.-Y.}\ \bibnamefont {Xie}}, \bibinfo {author} {\bibfnamefont {F.}~\bibnamefont {Meng}}, \bibinfo {author} {\bibfnamefont {Y.}~\bibnamefont {Liu}}, \bibinfo {author} {\bibfnamefont {Z.}~\bibnamefont {Bai}}, \bibinfo {author} {\bibfnamefont {S.}~\bibnamefont {Ji}}, \bibinfo {author} {\bibfnamefont {J.}~\bibnamefont {Zhang}}, \bibinfo {author} {\bibfnamefont {K.}~\bibnamefont {Jiang}}, \bibinfo {author} {\bibfnamefont {R.}~\bibnamefont {Zhong}}, \bibinfo {author} {\bibfnamefont {J.}~\bibnamefont {Schneeloch}}, \bibinfo {author} {\bibfnamefont {G.}~\bibnamefont {Gu}}, \bibinfo {author} {\bibfnamefont {L.}~\bibnamefont {Gu}}, \bibinfo {author} {\bibfnamefont {X.}~\bibnamefont {Ma}}, \bibinfo {author} {\bibfnamefont {D.}~\bibnamefont {Zhang}}, \ and\ \bibinfo {author} {\bibfnamefont {Q.-K.}\ \bibnamefont {Xue}},\ }\href {\doibase 10.1103/PhysRevX.11.031011} {\bibfield  {journal} {\bibinfo  {journal} {Phys. Rev. X}\ }\textbf {\bibinfo {volume} {11}},\ \bibinfo {pages} {031011} (\bibinfo {year} {2021})}\BibitemShut {NoStop}%
\bibitem [{\citenamefont {Lee}\ \emph {et~al.}(2021)\citenamefont {Lee}, \citenamefont {Lee}, \citenamefont {Kim}, \citenamefont {Choi}, \citenamefont {Park}, \citenamefont {Jang}, \citenamefont {Gu}, \citenamefont {Choi}, \citenamefont {Cho}, \citenamefont {Lee},\ and\ \citenamefont {Lee}}]{lee_twisted_2021}%
  \BibitemOpen
  \bibfield  {author} {\bibinfo {author} {\bibfnamefont {J.}~\bibnamefont {Lee}}, \bibinfo {author} {\bibfnamefont {W.}~\bibnamefont {Lee}}, \bibinfo {author} {\bibfnamefont {G.-Y.}\ \bibnamefont {Kim}}, \bibinfo {author} {\bibfnamefont {Y.-B.}\ \bibnamefont {Choi}}, \bibinfo {author} {\bibfnamefont {J.}~\bibnamefont {Park}}, \bibinfo {author} {\bibfnamefont {S.}~\bibnamefont {Jang}}, \bibinfo {author} {\bibfnamefont {G.}~\bibnamefont {Gu}}, \bibinfo {author} {\bibfnamefont {S.-Y.}\ \bibnamefont {Choi}}, \bibinfo {author} {\bibfnamefont {G.~Y.}\ \bibnamefont {Cho}}, \bibinfo {author} {\bibfnamefont {G.-H.}\ \bibnamefont {Lee}}, \ and\ \bibinfo {author} {\bibfnamefont {H.-J.}\ \bibnamefont {Lee}},\ }\href {\doibase 10.1021/acs.nanolett.1c03906} {\bibfield  {journal} {\bibinfo  {journal} {Nano Lett.}\ }\textbf {\bibinfo {volume} {21}},\ \bibinfo {pages} {10469} (\bibinfo {year} {2021})}\BibitemShut {NoStop}%
\bibitem [{\citenamefont {Zhao}\ \emph {et~al.}(2023)\citenamefont {Zhao},
  \citenamefont {Cui}, \citenamefont {Volkov}, \citenamefont {Yoo},
  \citenamefont {Lee}, \citenamefont {Gardener}, \citenamefont {Akey},
  \citenamefont {Engelke}, \citenamefont {Ronen}, \citenamefont {Zhong},
  \citenamefont {Gu}, \citenamefont {Plugge}, \citenamefont {Tummuru},
  \citenamefont {Kim}, \citenamefont {Franz}, \citenamefont {Pixley},
  \citenamefont {Poccia},\ and\ \citenamefont {Kim}}]{zhao_time-reversal_2023}%
  \BibitemOpen
  \bibfield  {author} {\bibinfo {author} {\bibfnamefont {S.~Y.~F.}\
  \bibnamefont {Zhao}}, \bibinfo {author} {\bibfnamefont {X.}~\bibnamefont
  {Cui}}, \bibinfo {author} {\bibfnamefont {P.~A.}\ \bibnamefont {Volkov}},
  \bibinfo {author} {\bibfnamefont {H.}~\bibnamefont {Yoo}}, \bibinfo {author}
  {\bibfnamefont {S.}~\bibnamefont {Lee}}, \bibinfo {author} {\bibfnamefont
  {J.~A.}\ \bibnamefont {Gardener}}, \bibinfo {author} {\bibfnamefont {A.~J.}\
  \bibnamefont {Akey}}, \bibinfo {author} {\bibfnamefont {R.}~\bibnamefont
  {Engelke}}, \bibinfo {author} {\bibfnamefont {Y.}~\bibnamefont {Ronen}},
  \bibinfo {author} {\bibfnamefont {R.}~\bibnamefont {Zhong}}, \bibinfo
  {author} {\bibfnamefont {G.}~\bibnamefont {Gu}}, \bibinfo {author}
  {\bibfnamefont {S.}~\bibnamefont {Plugge}}, \bibinfo {author} {\bibfnamefont
  {T.}~\bibnamefont {Tummuru}}, \bibinfo {author} {\bibfnamefont
  {M.}~\bibnamefont {Kim}}, \bibinfo {author} {\bibfnamefont {M.}~\bibnamefont
  {Franz}}, \bibinfo {author} {\bibfnamefont {J.~H.}\ \bibnamefont {Pixley}},
  \bibinfo {author} {\bibfnamefont {N.}~\bibnamefont {Poccia}}, \ and\ \bibinfo
  {author} {\bibfnamefont {P.}~\bibnamefont {Kim}},\ }\href {\doibase
  10.1126/science.abl8371} {\bibfield  {journal} {\bibinfo  {journal}
  {Science}\ ,\ \bibinfo {pages} {eabl8371}} (\bibinfo {year}
  {2023})}\BibitemShut {NoStop}%
\bibitem [{\citenamefont {Volkov}\ \emph {et~al.}(2021)\citenamefont {Volkov}, \citenamefont {Zhao}, \citenamefont {Poccia}, \citenamefont {Cui}, \citenamefont {Kim},\ and\ \citenamefont {Pixley}}]{volkov_josephson_2021}%
  \BibitemOpen
  \bibfield  {author} {\bibinfo {author} {\bibfnamefont {P.~A.}\ \bibnamefont {Volkov}}, \bibinfo {author} {\bibfnamefont {S.~Y.~F.}\ \bibnamefont {Zhao}}, \bibinfo {author} {\bibfnamefont {N.}~\bibnamefont {Poccia}}, \bibinfo {author} {\bibfnamefont {X.}~\bibnamefont {Cui}}, \bibinfo {author} {\bibfnamefont {P.}~\bibnamefont {Kim}}, \ and\ \bibinfo {author} {\bibfnamefont {J.~H.}\ \bibnamefont {Pixley}},\ }\href {http://arxiv.org/abs/2108.13456} {\enquote {\bibinfo {title} {Josephson effects in twisted nodal superconductors},}\ } (\bibinfo {year} {2021}),\ \bibinfo {note} {arXiv:2108.13456 [cond-mat]}\BibitemShut {NoStop}%
\bibitem [{\citenamefont {Tummuru}\ \emph {et~al.}(2022{\natexlab{b}})\citenamefont {Tummuru}, \citenamefont {Plugge},\ and\ \citenamefont {Franz}}]{tummuru_josephson_2022}%
  \BibitemOpen
  \bibfield  {author} {\bibinfo {author} {\bibfnamefont {T.}~\bibnamefont {Tummuru}}, \bibinfo {author} {\bibfnamefont {S.}~\bibnamefont {Plugge}}, \ and\ \bibinfo {author} {\bibfnamefont {M.}~\bibnamefont {Franz}},\ }\href {\doibase 10.1103/PhysRevB.105.064501} {\bibfield  {journal} {\bibinfo  {journal} {Phys. Rev. B}\ }\textbf {\bibinfo {volume} {105}},\ \bibinfo {pages} {064501} (\bibinfo {year} {2022}{\natexlab{b}})}\BibitemShut {NoStop}%
\bibitem [{\citenamefont {Dash}\ and\ \citenamefont {Sénéchal}(2019)}]{dash_pseudogap_2019}%
  \BibitemOpen
  \bibfield  {author} {\bibinfo {author} {\bibfnamefont {S.~S.}\ \bibnamefont {Dash}}\ and\ \bibinfo {author} {\bibfnamefont {D.}~\bibnamefont {Sénéchal}},\ }\href {\doibase 10.1103/PhysRevB.100.214509} {\bibfield  {journal} {\bibinfo  {journal} {Phys. Rev. B}\ }\textbf {\bibinfo {volume} {100}},\ \bibinfo {pages} {214509} (\bibinfo {year} {2019})}\BibitemShut {NoStop}%
\bibitem [{\citenamefont {Song}\ \emph {et~al.}(2022)\citenamefont {Song}, \citenamefont {Zhang},\ and\ \citenamefont {Vishwanath}}]{song_doping_2022}%
  \BibitemOpen
  \bibfield  {author} {\bibinfo {author} {\bibfnamefont {X.-Y.}\ \bibnamefont {Song}}, \bibinfo {author} {\bibfnamefont {Y.-H.}\ \bibnamefont {Zhang}}, \ and\ \bibinfo {author} {\bibfnamefont {A.}~\bibnamefont {Vishwanath}},\ }\href {\doibase 10.1103/PhysRevB.105.L201102} {\bibfield  {journal} {\bibinfo  {journal} {Phys. Rev. B}\ }\textbf {\bibinfo {volume} {105}},\ \bibinfo {pages} {L201102} (\bibinfo {year} {2022})},\ \bibinfo {note} {publisher: American Physical Society}\BibitemShut {NoStop}%
\bibitem [{\citenamefont {Sénéchal}\ \emph {et~al.}(2013)\citenamefont {Sénéchal}, \citenamefont {Day}, \citenamefont {Bouliane},\ and\ \citenamefont {Tremblay}}]{senechal_resilience_2013}%
  \BibitemOpen
  \bibfield  {author} {\bibinfo {author} {\bibfnamefont {D.}~\bibnamefont {Sénéchal}}, \bibinfo {author} {\bibfnamefont {A.~G.~R.}\ \bibnamefont {Day}}, \bibinfo {author} {\bibfnamefont {V.}~\bibnamefont {Bouliane}}, \ and\ \bibinfo {author} {\bibfnamefont {A.-M.~S.}\ \bibnamefont {Tremblay}},\ }\href {\doibase 10.1103/PhysRevB.87.075123} {\bibfield  {journal} {\bibinfo  {journal} {Phys. Rev. B}\ }\textbf {\bibinfo {volume} {87}},\ \bibinfo {pages} {075123} (\bibinfo {year} {2013})}\BibitemShut {NoStop}%
\bibitem [{\citenamefont {Markiewicz}\ \emph {et~al.}(2005)\citenamefont {Markiewicz}, \citenamefont {Sahrakorpi}, \citenamefont {Lindroos}, \citenamefont {Lin},\ and\ \citenamefont {Bansil}}]{markiewicz_one-band_2005}%
  \BibitemOpen
  \bibfield  {author} {\bibinfo {author} {\bibfnamefont {R.~S.}\ \bibnamefont {Markiewicz}}, \bibinfo {author} {\bibfnamefont {S.}~\bibnamefont {Sahrakorpi}}, \bibinfo {author} {\bibfnamefont {M.}~\bibnamefont {Lindroos}}, \bibinfo {author} {\bibfnamefont {H.}~\bibnamefont {Lin}}, \ and\ \bibinfo {author} {\bibfnamefont {A.}~\bibnamefont {Bansil}},\ }\href {\doibase 10.1103/PhysRevB.72.054519} {\bibfield  {journal} {\bibinfo  {journal} {Phys. Rev. B}\ }\textbf {\bibinfo {volume} {72}},\ \bibinfo {pages} {054519} (\bibinfo {year} {2005})}\BibitemShut {NoStop}%
\bibitem [{\citenamefont {Potthoff}\ \emph {et~al.}(2003)\citenamefont {Potthoff}, \citenamefont {Aichhorn},\ and\ \citenamefont {Dahnken}}]{potthoff_variational_2003}%
  \BibitemOpen
  \bibfield  {author} {\bibinfo {author} {\bibfnamefont {M.}~\bibnamefont {Potthoff}}, \bibinfo {author} {\bibfnamefont {M.}~\bibnamefont {Aichhorn}}, \ and\ \bibinfo {author} {\bibfnamefont {C.}~\bibnamefont {Dahnken}},\ }\href {\doibase 10.1103/PhysRevLett.91.206402} {\bibfield  {journal} {\bibinfo  {journal} {Phys. Rev. Lett.}\ }\textbf {\bibinfo {volume} {91}},\ \bibinfo {pages} {206402} (\bibinfo {year} {2003})}\BibitemShut {NoStop}%
\bibitem [{\citenamefont {Dahnken}\ \emph {et~al.}(2004)\citenamefont {Dahnken}, \citenamefont {Aichhorn}, \citenamefont {Hanke}, \citenamefont {Arrigoni},\ and\ \citenamefont {Potthoff}}]{dahnken_variational_2004}%
  \BibitemOpen
  \bibfield  {author} {\bibinfo {author} {\bibfnamefont {C.}~\bibnamefont {Dahnken}}, \bibinfo {author} {\bibfnamefont {M.}~\bibnamefont {Aichhorn}}, \bibinfo {author} {\bibfnamefont {W.}~\bibnamefont {Hanke}}, \bibinfo {author} {\bibfnamefont {E.}~\bibnamefont {Arrigoni}}, \ and\ \bibinfo {author} {\bibfnamefont {M.}~\bibnamefont {Potthoff}},\ }\href {\doibase 10.1103/PhysRevB.70.245110} {\bibfield  {journal} {\bibinfo  {journal} {Phys. Rev. B}\ }\textbf {\bibinfo {volume} {70}},\ \bibinfo {pages} {245110} (\bibinfo {year} {2004})}\BibitemShut {NoStop}%
\bibitem [{\citenamefont {Sahebsara}\ and\ \citenamefont {Sénéchal}(2008)}]{sahebsara_hubbard_2008}%
  \BibitemOpen
  \bibfield  {author} {\bibinfo {author} {\bibfnamefont {P.}~\bibnamefont {Sahebsara}}\ and\ \bibinfo {author} {\bibfnamefont {D.}~\bibnamefont {Sénéchal}},\ }\href {\doibase 10.1103/PhysRevLett.100.136402} {\bibfield  {journal} {\bibinfo  {journal} {Phys. Rev. Lett.}\ }\textbf {\bibinfo {volume} {100}},\ \bibinfo {pages} {136402} (\bibinfo {year} {2008})}\BibitemShut {NoStop}%
\bibitem [{\citenamefont {Laubach}\ \emph {et~al.}(2014)\citenamefont {Laubach}, \citenamefont {Reuther}, \citenamefont {Thomale},\ and\ \citenamefont {Rachel}}]{laubach2014}%
  \BibitemOpen
  \bibfield  {author} {\bibinfo {author} {\bibfnamefont {M.}~\bibnamefont {Laubach}}, \bibinfo {author} {\bibfnamefont {J.}~\bibnamefont {Reuther}}, \bibinfo {author} {\bibfnamefont {R.}~\bibnamefont {Thomale}}, \ and\ \bibinfo {author} {\bibfnamefont {S.}~\bibnamefont {Rachel}},\ }\href@noop {} {\bibfield  {journal} {\bibinfo  {journal} {Phys. Rev. B}\ }\textbf {\bibinfo {volume} {90}},\ \bibinfo {pages} {165136} (\bibinfo {year} {2014})}\BibitemShut {NoStop}%
\bibitem [{\citenamefont {S{\'e}n{\'e}chal}\ \emph {et~al.}(2005)\citenamefont {S{\'e}n{\'e}chal}, \citenamefont {Lavertu}, \citenamefont {a.~Marois},\ and\ \citenamefont {Tremblay}}]{senechal2005}%
  \BibitemOpen
  \bibfield  {author} {\bibinfo {author} {\bibfnamefont {D.}~\bibnamefont {S{\'e}n{\'e}chal}}, \bibinfo {author} {\bibfnamefont {P.~L.}\ \bibnamefont {Lavertu}}, \bibinfo {author} {\bibfnamefont {M.}~\bibnamefont {a.~Marois}}, \ and\ \bibinfo {author} {\bibfnamefont {{\relax AM}.}~\bibnamefont {Tremblay}},\ }\href {\doibase 10.1103/PhysRevLett.94.156404} {\bibfield  {journal} {\bibinfo  {journal} {Physical Review Letters}\ }\textbf {\bibinfo {volume} {94}},\ \bibinfo {pages} {156404} (\bibinfo {year} {2005})}\BibitemShut {NoStop}%
\bibitem [{\citenamefont {Faye}\ and\ \citenamefont {Sénéchal}(2017)}]{faye_interplay_2017}%
  \BibitemOpen
  \bibfield  {author} {\bibinfo {author} {\bibfnamefont {J.~P.~L.}\ \bibnamefont {Faye}}\ and\ \bibinfo {author} {\bibfnamefont {D.}~\bibnamefont {Sénéchal}},\ }\href {\doibase 10.1103/PhysRevB.95.115127} {\bibfield  {journal} {\bibinfo  {journal} {Phys. Rev. B}\ }\textbf {\bibinfo {volume} {95}},\ \bibinfo {pages} {115127} (\bibinfo {year} {2017})}\BibitemShut {NoStop}%
\bibitem [{\citenamefont {Potthoff}(2012)}]{potthoff_variational_2012}%
  \BibitemOpen
  \bibfield  {author} {\bibinfo {author} {\bibfnamefont {M.}~\bibnamefont {Potthoff}},\ }in\ \href@noop {} {\emph {\bibinfo {booktitle} {Strongly {Correlated} {Systems}: {Theoretical} {Methods}}}},\ Vol.\ \bibinfo {volume} {171},\ \bibinfo {editor} {edited by\ \bibinfo {editor} {\bibfnamefont {A.}~\bibnamefont {Avella}}\ and\ \bibinfo {editor} {\bibfnamefont {F.}~\bibnamefont {Mancini}}}\ (\bibinfo  {publisher} {Springer Berlin Heidelberg},\ \bibinfo {address} {Berlin, Heidelberg},\ \bibinfo {year} {2012})\ pp.\ \bibinfo {pages} {303--339}\BibitemShut {NoStop}%
 \bibitem [{\citenamefont {Potthoff}(2014)}]{potthoff_dmft_2014}%
  \BibitemOpen
  \bibfield  {author} {\bibinfo {author} {\bibfnamefont {M.}~\bibnamefont
  {Potthoff}},\ }\href@noop {} {{\emph {\bibinfo {title}
  {{DMFT} at 25: infinite dimensions: lecture notes of the {Autumn} {School} on
  {Correlated} {Electrons} 2014}}}},\ edited by\ \bibinfo {editor}
  {\bibfnamefont {E.}~\bibnamefont {Pavarini}}, \bibinfo {editor}
  {\bibfnamefont {E.}~\bibnamefont {Koch}}, \bibinfo {editor} {\bibfnamefont
  {D.}~\bibnamefont {Vollhardt}}, \ and\ \bibinfo {editor} {\bibfnamefont
  {A.~I.}\ \bibnamefont {Lichtenstein}},\ \bibinfo {series} {Schriften des
  {Forschungszentrums} {Jülich}. {Reihe} {Modeling} and {Simulation}}\ No.\
  \bibinfo {number} {Band 4}\ (\bibinfo  {publisher} {Forschungszentrum
  Jülich, Zentralbibliothek, Verl},\ \bibinfo {address} {Jülich},\ \bibinfo
  {year} {2014})\BibitemShut {NoStop}%
\bibitem [{\citenamefont {Maier}\ \emph {et~al.}(2005)\citenamefont {Maier}, \citenamefont {Jarrell}, \citenamefont {Schulthess}, \citenamefont {Kent},\ and\ \citenamefont {White}}]{maier_systematic_2005}%
  \BibitemOpen
  \bibfield  {author} {\bibinfo {author} {\bibfnamefont {T.~A.}\ \bibnamefont {Maier}}, \bibinfo {author} {\bibfnamefont {M.}~\bibnamefont {Jarrell}}, \bibinfo {author} {\bibfnamefont {T.~C.}\ \bibnamefont {Schulthess}}, \bibinfo {author} {\bibfnamefont {P.~R.~C.}\ \bibnamefont {Kent}}, \ and\ \bibinfo {author} {\bibfnamefont {J.~B.}\ \bibnamefont {White}},\ }\href {\doibase 10.1103/PhysRevLett.95.237001} {\bibfield  {journal} {\bibinfo  {journal} {Phys. Rev. Lett.}\ }\textbf {\bibinfo {volume} {95}},\ \bibinfo {pages} {237001} (\bibinfo {year} {2005})}\BibitemShut {NoStop}%
\bibitem [{\citenamefont {Kaba}\ and\ \citenamefont {Sénéchal}(2019)}]{kaba_group-theoretical_2019}%
  \BibitemOpen
  \bibfield  {author} {\bibinfo {author} {\bibfnamefont {S.-O.}\ \bibnamefont {Kaba}}\ and\ \bibinfo {author} {\bibfnamefont {D.}~\bibnamefont {Sénéchal}},\ }\href {\doibase 10.1103/PhysRevB.100.214507} {\bibfield  {journal} {\bibinfo  {journal} {Phys. Rev. B}\ }\textbf {\bibinfo {volume} {100}},\ \bibinfo {pages} {214507} (\bibinfo {year} {2019})}\BibitemShut {NoStop}%
\bibitem [{\citenamefont {Yuan}\ \emph {et~al.}(2023)\citenamefont {Yuan}, \citenamefont {Vituri}, \citenamefont {Berg}, \citenamefont {Spivak},\ and\ \citenamefont {Kivelson}}]{yuan_inhomogeneity-induced_2023}%
  \BibitemOpen
  \bibfield  {author} {\bibinfo {author} {\bibfnamefont {A.~C.}\ \bibnamefont {Yuan}}, \bibinfo {author} {\bibfnamefont {Y.}~\bibnamefont {Vituri}}, \bibinfo {author} {\bibfnamefont {E.}~\bibnamefont {Berg}}, \bibinfo {author} {\bibfnamefont {B.}~\bibnamefont {Spivak}}, \ and\ \bibinfo {author} {\bibfnamefont {S.~A.}\ \bibnamefont {Kivelson}},\ }\href {http://arxiv.org/abs/2305.15472} {\enquote {\bibinfo {title} {Inhomogeneity-{Induced} {Time}-{Reversal} {Symmetry} {Breaking} in {Cuprate} {Twist}-{Junctions}},}\ } (\bibinfo {year} {2023}),\ \bibinfo {note} {arXiv:2305.15472 [cond-mat]}\BibitemShut {NoStop}%
\end{thebibliography}
% \end{document}


\end{document}